\documentclass{aa}  

\usepackage{graphicx}
\usepackage{txfonts}
\usepackage[breaklinks=True]{hyperref}
\begin{document}

   \title{The first and second data releases of the Kilo-Degree Survey}

   \author{Jelte T. A. de Jong\inst{1}
          \and Gijs A. Verdoes Kleijn\inst{2}
          \and Danny R. Boxhoorn\inst{2}
          \and Hugo Buddelmeijer\inst{2}
          \and Massimo Capaccioli\inst{3}
          \and Fedor Getman\inst{3}
          \and Aniello Grado\inst{3}
          \and Ewout Helmich\inst{1}
          \and Zhuoyi Huang\inst{3}
          \and Nancy Irisarri\inst{1}
          \and Konrad Kuijken\inst{1}
          \and Francesco La Barbera\inst{3}
          \and John P. McFarland\inst{2}
          \and Nicola R. Napolitano\inst{3}
          \and Mario Radovich\inst{4}
          \and Gert Sikkema\inst{2}
          \and Edwin A. Valentijn\inst{2}
          \and Kor G. Begeman\inst{2}
          \and Massimo Brescia\inst{3}
          \and Stefano Cavuoti\inst{3}
          \and Ami Choi\inst{5}
          \and Oliver-Mark Cordes\inst{6}
          \and Giovanni Covone\inst{7}
          \and Massimo Dall'Ora\inst{3}
          \and Hendrik Hildebrandt\inst{6}
          \and Giuseppe Longo\inst{7}
          \and Reiko Nakajima\inst{6}
          \and Maurizio Paolillo\inst{7,8}
          \and Emanuella Puddu\inst{3}
          \and Agatino Rifatto\inst{3}
          \and Crescenzo Tortora\inst{3}
          \and Edo van Uitert\inst{6,9}
          \and Axel Buddendiek\inst{6}
          \and Joachim Harnois-D\'eraps\inst{10}
          \and Thomas Erben\inst{6}
          \and Martin B. Eriksen\inst{1}
          \and Catherine Heymans\inst{5}
          \and Henk Hoekstra\inst{1}
          \and Benjamin Joachimi\inst{9}
          \and Thomas D. Kitching\inst{11}
          \and Dominik Klaes\inst{6}
          \and L\'eon V. E. Koopmans\inst{2}
          \and Fabian K\"ohlinger\inst{1}
          \and Nivya Roy\inst{7}
          \and Crist\'obal Sif\'on\inst{1}
          \and Peter Schneider\inst{6}
          \and Will J. Sutherland\inst{12}
          \and Massimo Viola\inst{1}
          \and Willem-Jan Vriend\inst{2}
          }

   \institute{Leiden Observatory, Leiden University, P.O. Box 9513, 2300 RA Leiden, the Netherlands\\
              \email{jdejong@strw.leidenuniv.nl}
         \and
              Kapteyn Astronomical Institute, University of Groningen, P.O. Box 800, 9700 AV Groningen, the Netherlands
         \and
              INAF - Osservatorio Astronomico di Capodimonte, Via Moiariello 16 -80131 Napoli, Italy
         \and
              INAF - Osservatorio Astronomico di Padova, via dell'Osservatorio 5, 35122 Padova, Italy
         \and
              Scottish Universities Physics Alliance, Institute for Astronomy, University of Edinburgh, Royal Observatory, Blackford Hill, Edinburgh EH9 3HJ, United Kingdom
         \and
              Argelander-Institut f\"ur Astronomie, Auf dem H\"ugel 71, D-53121 Bonn, Germany
         \and
              Department of Physics, University Federico II, Via Cinthia 6, 80126 Napoli, Italy
         \and
              Agenzia Spaziale Italiana - Science Data Center, Via del Politecnico snc, 00133 Roma, Italy
         \and
              Department of Physics \& Astronomy, University College London, Gower Street, London WC1E 6BT, United Kingdom
         \and
              Department of Physics and Astronomy, University of British Columbia, BC V6T 1Z1, Canada
         \and
              Mullard Space Science Laboratory, University College London, Holmbury St Mary, Dorking, Surrey RH5 6NT, United Kingdom
         \and
              School of Physics and Astronomy, Queen Mary University of London, Mile End Road, London E1 4NS, United Kingdom 
             }

   \date{}

 
  \abstract
   {The Kilo-Degree Survey (KiDS) is an optical wide-field imaging survey carried out with the VLT Survey Telescope and the OmegaCAM camera. KiDS will image 1500 square degrees in four filters ($ugri$), and together with its near-infrared counterpart VIKING will produce deep photometry in nine bands. Designed for weak lensing shape and photometric redshift measurements, its core science driver is mapping the large-scale matter distribution in the Universe back to a redshift of $\sim$0.5. Secondary science cases include galaxy evolution, Milky Way structure, and the detection of high-redshift clusters and quasars.}
   {KiDS is an ESO Public Survey and dedicated to serving the astronomical community with high-quality data products derived from the survey data. Public data releases, the first two of which are presented here, are crucial for enabling independent confirmation of the survey's scientific value. The achieved data quality and initial scientific utilization are reviewed in order to validate the survey data.}
   {A dedicated pipeline and data management system based on {\sc Astro-WISE}, combined with newly developed masking and source classification tools, is used for the production of the data products described here. Science projects based on these data products and preliminary results are outlined.}
   {For 148 survey tiles ($\approx$160 sq.deg.) stacked $ugri$ images have been released, accompanied by weight maps, masks, source lists, and a multi-band source catalogue. Limiting magnitudes are typically 24.3, 25.1, 24.9, 23.8 (5$\sigma$ in a 2\arcsec aperture) in $ugri$, respectively, and the typical $r$-band PSF size is less than 0.7\arcsec. The photometry prior to global homogenization is stable at the $\sim$2\% (4\%) level in $gri$ ($u$) with some outliers due to non-photometric conditions, while the astrometry shows a typical 2-D RMS of 0.03\arcsec. Early scientific results include the detection of nine high-z QSOs, fifteen candidate strong gravitational lenses, high-quality photometric redshifts and structural parameters for hundreds of thousands of galaxies.}
   {}


   \maketitle

\titlerunning{KiDS data release 1 and 2}
\authorrunning{J. T. A. de Jong et al.}

\section{Introduction}
\label{Sec:Introduction}

Sensitive, wide-field astronomical surveys have proven to be extremely useful scientific resources, not only for the specific scientific use cases they are designed for, but also due to their legacy value and as source-finders for the largest telescopes. With the arrival of VISTA in 2010 and the VLT Survey Telescope (VST) in 2011, the European astronomical community now has access to dedicated survey telescopes both in the infrared and the optical. During the first years of operations the majority of observing time on both telescopes is dedicated to a number of public surveys, large observational programs selected by ESO that serve the astronomical community with regular public data releases\footnote{\url{http://www.eso.org/sci/observing/PublicSurveys.html}}. 
On VST these are the VST Photometric H-$\alpha$ Survey of the Southern Galactic Plane \citep[VPHAS+, ][]{vphas+}, VST ATLAS \citep{vstatlas}, which covers 5000 square degrees in the Southern Galactic Cap to similar depth as the Sloan Digital Sky Survey \citep[SDSS, ][]{sdssdr9}, and the Kilo-Degree Survey \citep[KiDS, ][]{kids_expast}.
The Kilo-Degree Survey (KiDS) is the largest public survey on the VST in terms of observing time and will image 1500 square degrees of extragalactic sky in four filters, $u$, $g$, $r$, and $i$. Together with the VISTA Kilo-Degree Infrared Galaxy Survey \citep[VIKING, ][]{viking}, a sister survey on VISTA that observes the same area in $Z$, $Y$, $J$, $H$ and $K_S$, this will result in a deep nine-band data set with, in total, 185 minutes of observing time per survey tile over all filters. Part of a long heritage of ever improving wide-field optical sky surveys, the combination of superb image quality with wide wavelength coverage provided by KiDS/VIKING will be unique for surveys of this size and depth for the foreseeable future.

The main science driver for KiDS is mapping of the large-scale mass distribution in the Universe and constraining its expansion history by means of weak gravitational lensing and photometric redshift measurements. These goals put stringent requirements on image quality and stability, photometric depth and calibration accuracy. However, this high data quality will be beneficial for many other science cases, and secondary science drivers for KiDS include a varied set of topics, such as the evolution of clusters and galaxies, galaxy scaling relations, Galactic stellar halo structure, and searches for rare objects such as strong gravitational lenses and high-redshift quasars. How KiDS will benefit the primary and secondary science drivers is discussed in more detail in \cite{kids_expast}.

KiDS data will be released to the astronomical community in yearly Public survey releases of those survey tiles that have been observed in all four filters. In this paper we describe the data products of the first two such KiDS data releases (DR), covering a total of 148 survey tiles. Apart from the released data products (Sect. \ref{Sec:DataDistribution}), observational set-up (Sect. \ref{Sec:ObservationalSetup}), the data processing pipeline (Sect. \ref{Sec:DataProcessing}) and data quality (Sect. \ref{Sec:DataQuality}), we also review the scientific research that is currently being carried out by the KiDS team using the data presented here in Sect. \ref{Sec:ScienceResults}. Finally, the paper is closed with a brief summary and outlook in Sect. \ref{Sec:Summary}.

   \begin{figure*}
   \centering
   \includegraphics[width=\textwidth]{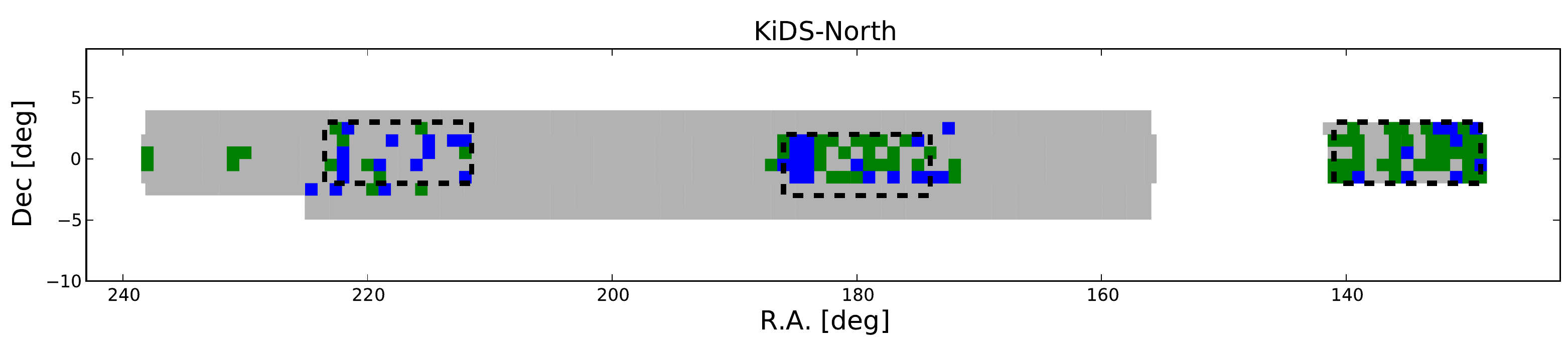}
   \includegraphics[width=\textwidth]{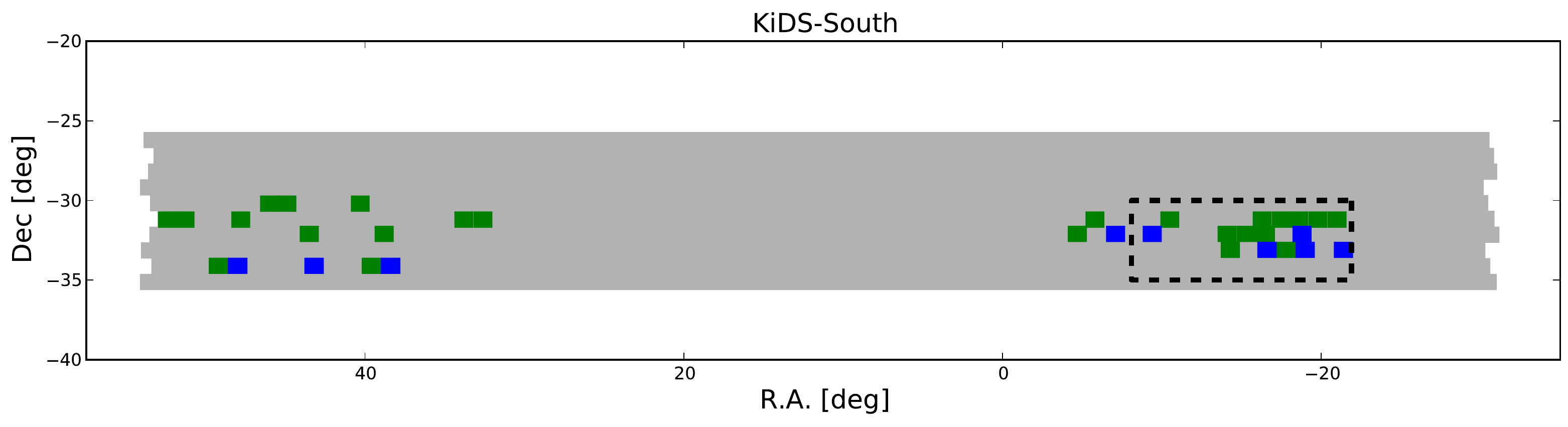}
   \caption{Sky distribution of survey tiles released in KiDS-ESO-DR1 (blue) and KiDS-ESO-DR2 (green), with the full KiDS area is shown in grey. {\it Top:} KiDS-North. {\it Bottom:} KiDS- South. The multi-band source catalogue covers the combined area (blue + green). The black dashed lines delineate the locations of the GAMA fields.}
              \label{Fig:SkyDistribution}
    \end{figure*}

\section{Data description}
\label{Sec:DataDistribution}

\subsection{General properties}
\label{Sec:GeneralProperties}

KiDS is designed for weak lensing shape and photometric redshift measurements, with weak gravitational lensing
tomography as its ultimate science goal. This goal is efficiently reached at the VST with
deep, good-seeing $r$-band observations taken under the best
atmospheric conditions, supplemented with somewhat shallower and less
sharp $u,g,i$ images. Galaxy shapes for the lensing studies can
then be measured on the $r$ band data, while the other bands yield colours that can be used to
derive photometric redshifts for these same galaxies. The detailed
implementation of these requirements is described below in
Sect. \ref{Sec:ObservationalSetup}; typical values for the resolution (PSF
FWHM) and limiting magnitude (5-$\sigma$ AB magnitude in a 2\arcsec
aperture) are (1.0\arcsec, 0.8\arcsec, 0.65\arcsec, 0.85\arcsec) and (24.3, 25.1, 24.9,
23.8) in $u$, $g$, $r$ and $i$ bands, respectively (see Sect. \ref{Sec:DataQuality} for a detailed description). The $i$ band
data are significantly less uniform in quality than the other bands
because they are mostly taken when the moon is above the
horizon.

\subsection{OmegaCAM and VST}

Designed to provide superb and uniform image quality over the full field-of-view (FOV), the combination of VST and the OmegaCAM wide-field imager is ideal for a survey such as KiDS. The VST, an alt-az mounted modified Ritchey-Cretien telescope, uses a two-lens field corrector and has active primary and secondary mirrors. OmegaCAM is a 268 Megapixel CCD imaging camera with a one square degree FOV. Located at the Cassegrain focus, it is the only instrument on the VST. The filters and dewar window are part of the total optical design, with the latter acting as a field lens. The science array consists of 32 thinned, low-noise 2k$\times$4k E2V devices. With little aberration over the full FOV the 15$\mu$ pixel size translates to a constant pixel scale of 0.21\arcsec\, with small gaps of 25\arcsec and 85\arcsec between the CCDs. Wavefront sensing, using two auxiliary CCDs on either side of the science array, is employed to monitor and optimize the optical set-up in real time. Auto-guiding of tracking and rotation makes use of two more auxiliary CCDs. Combined, the VST and OmegaCAM provide a PSF equal to the atmospheric seeing over the full FOV down to 0.6\arcsec. More details on the VST and OmegaCAM can be found in \cite{vst} and \cite{omegacam}, respectively.

\subsection{Sky distribution}
\label{Sec:SkyDistribution}

KiDS data releases consist of $\sim$1 square degree tiles that have
been successfully observed in all four survey filters
($u$,$g$,$r$,$i$). To maximize the synergy with existing spectroscopic
surveys and between KiDS and VIKING early on, the latter surveys both
prioritize the observations in the Galaxy And Mass Assembly
\citep[GAMA, ][]{gama} fields G9, G12, G15 and G23.

The first public data release of KiDS (KiDS-ESO-DR1) was issued in
July 2013 and contains imaging data, masks and single-band source
lists for all tiles observed in all four filters during the first year
of regular operations (15 October 2011 to 31 September 2012),
including data taken during Early Science Time (EST, 13 August to 15
October 2011), a total of 50 tiles.  The second data release
(KiDS-ESO-DR2) was available in February 2015 and contains the
same data products for all tiles for which observations were completed
during the second year of regular operations (1 October 2012 to 31
September 2013), a total of 98 tiles. Since the processing pipeline
used for KiDS-ESO-DR2 is practically identical to that used for
KiDS-ESO-DR1, the tiles released in the former complement the data set
of the latter, making KiDS-ESO-DR2 an incremental release. Apart from
the data products mentioned above, KiDS-ESO-DR2 also provides a
multi-band source catalogue based on the combined set of 148 tiles
released in the first two data releases.

Many other fields have been observed in a subset of the filters and
will be included in future releases once their wavelength coverage is
complete.  Figure \ref{Fig:SkyDistribution} shows the sky distribution
of the tiles included in KiDS-ESO-DR1/2 within the KiDS
fields. A complete list of all tiles with data quality parameters can
be found on the KiDS website\footnote{\url{http://kids.strw.leidenuniv.nl/DR2}}.

\subsection{Data products}
\label{Sec:DataProducts}

For every tile in the first two data releases of KiDS the following data products are
included for each of the bands, $u$, $g$, $r$, and $i$:
\begin{itemize}
  \item astrometrically and photometrically calibrated, stacked images (``coadds'')
  \item weight maps
  \item flag maps (``masks'') that flag saturated pixels, reflection halos, read-out spikes, etc.
  \item single-band source lists
\end{itemize}

The multi-band source catalogue encompassing the combined area of the
two data releases, although forming one large catalogue, is stored in
multiple files, namely one per survey tile.

\subsubsection{Coadded image units and gain}
\label{Sec:coaddunits}

The final calibrated, coadded images have a uniform pixel scale of 0.2
arcsec. The pixel units are fluxes relative to the flux corresponding
to magnitude = 0. This means that the magnitude $m$ corresponding to a pixel value $f$ is given by:
\begin{equation}
\label{Eq:fluxscale}
m = -2.5 \log_{10} f.
\end{equation}

The gain varies slightly over the field-of-view because of the photometric homogenization procedure described in Sect. \ref{Sec:PhotometricCalibration}. An average effective gain is provided online\footnote{\url{http://kids.strw.leidenuniv.nl/DR2/data_table.php}}. An example of a FITS header of a coadded image is also provided online\footnote{\url{http://kids.strw.leidenuniv.nl/DR2/example_imageheader.txt}}.

\subsubsection{Single-band source list contents}

For each tile single-band source lists are provided for each of the
survey filters. To increase the usefulness and versatility of these
source lists, an extensive set of magnitude and shape parameters are
included, including a large number (27) of aperture magnitudes. The
latter allows users to use interpolation methods (e.g. ``curve of
growth'') to derive their own aperture corrections or total
magnitudes. Also provided is a star-galaxy separation parameter and
information on the mask regions (see Sect. \ref{Sec:Masking}) that
might affect individual source measurements. Details on the production of these source lists, including source detection and other measurements are discussed in Sect. 
\ref{Sec:SingleBandSourceLists}.

Table \ref{Tab:singlebandcolumns} lists the columns that are present
in the single-band source lists provided in KiDS-ESO-DR1/2. Note that 
of the 27 aperture flux columns only the ones for the smallest aperture (2 pixels, or
0.4\arcsec\ diameter) and the largest aperture (200 pixels, or
40\arcsec\ diameter) are listed.

\subsubsection{Multi-band catalogue contents}

KiDS-ESO-DR2 includes a multi-band source catalogue. This catalogue is
based on source detection in the $r$-band images. While magnitudes are
measured for all of the filters, the star-galaxy separation,
positional, and shape parameters are based on the $r$-band data. The
choice of $r$-band is motivated by the fact that it typically has the best
image quality and thus provides the most reliable source positions and
shapes. Seeing differences between observations in the different
filters are mitigated by the inclusion of aperture corrections. Details on the production of this catalogue are discussed Sect. \ref{Sec:MultiBandCatalogue} and Table
\ref{Tab:MultiBandColumns} lists the columns present in the multi-band
source lists in KiDS-ESO-DR2.

   \begin{figure}
   \centering
   \includegraphics[width=9cm]{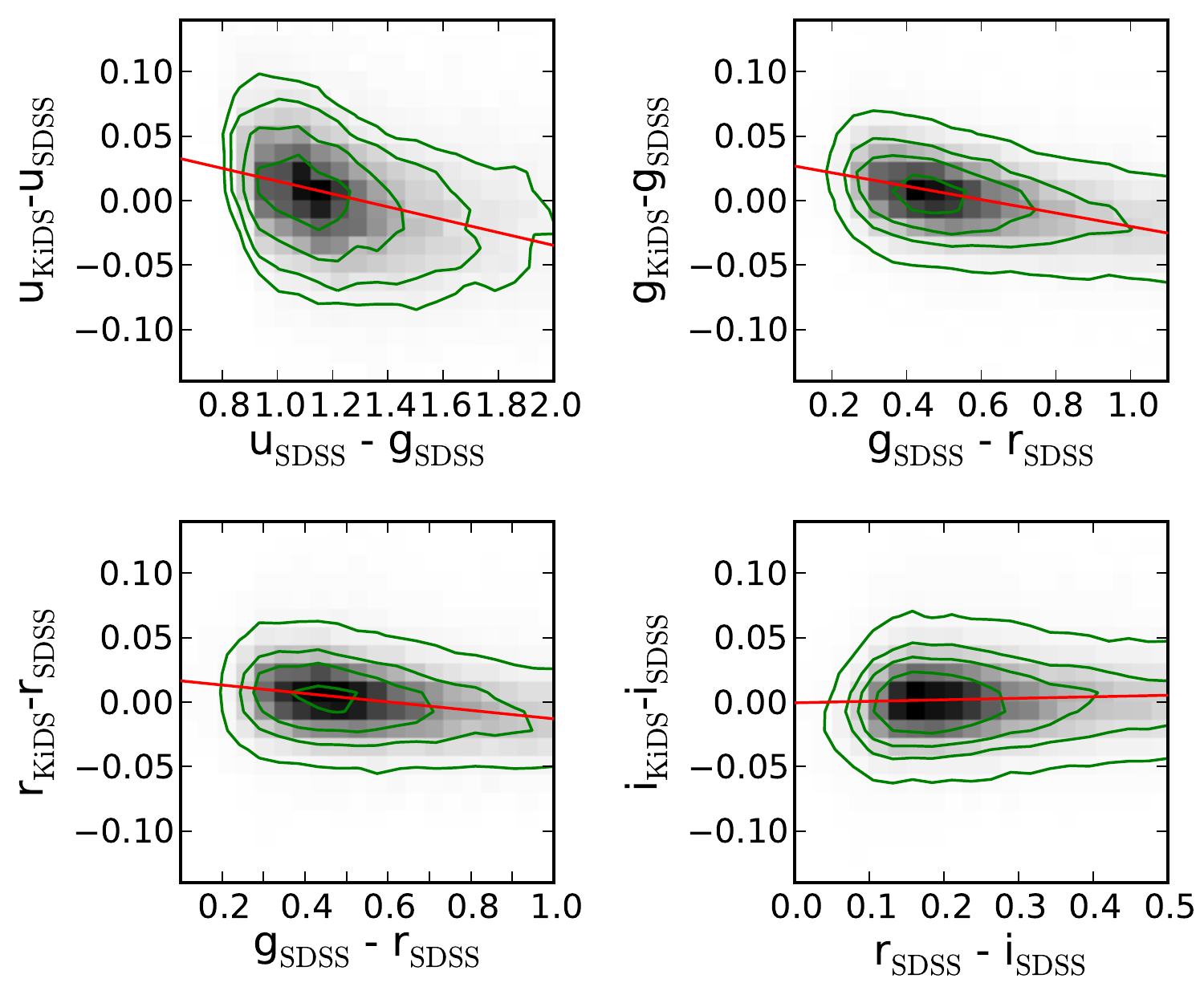}
   \caption{Derivation of colour-terms of KiDS photometry with respect to the SDSS photometric system. 
The distribution of stars is shown as a 2-D histogram, overlayed with isodensity contours (green) and fit by a linear relation (red line). For $u$ the contours correspond to 50, 100, 200 and 350 stars per bin, and for $g$, $r$ and $i$ to 100, 350, 600 and 1000 stars per bin. The absolute calibration between KiDS and SDSS (horizontal offset) is arbitrary.}
    \label{Fig:colourTerms}
    \end{figure}

\subsection{Colour terms}
\label{Sec:colourTerms}

The photometric calibration provided in KiDS-ESO-DR1 and KiDS-ESO-DR2
is in AB magnitudes in the instrumental system. Colour-terms have been
calculated with respect to the SDSS photometric system.  

Aperture-corrected magnitudes taken from the multi-band catalogue
were matched to SDSS DR8 \citep{aihara11} PSF magnitudes of point-like sources. For
each filter, the median offset to SDSS is first subtracted, rejecting
tiles where this offset exceeds 0.1 mag in any of the
bands (10 tiles), in order to prevent poor photometric calibration to affect the results. 
The fit is performed on all points from the remaining tiles. Figure
\ref{Fig:colourTerms} shows the photometric comparison and the
following equations give the resulting colour terms:

\begin{eqnarray}
u_{\rm{KiDS}} - u_{\rm{SDSS}} = (-0.050 \pm 0.002)(u_{\rm{SDSS}} - g_{\rm{SDSS}}),\\
g_{\rm{KiDS}} - g_{\rm{SDSS}} = (-0.052 \pm 0.002)(g_{\rm{SDSS}} - r_{\rm{SDSS}}),\\
r_{\rm{KiDS}} - r_{\rm{SDSS}} = (-0.033 \pm 0.002) (g_{\rm{SDSS}} - r_{\rm{SDSS}}),\\
i_{\rm{KiDS}} - i_{\rm{SDSS}} = (+0.012 \pm 0.002)(r_{\rm{SDSS}} - i_{\rm{SDSS}}).
\end{eqnarray}

\subsection{Data access}
\label{Sec:DataAccess}

Data from the first two KiDS data releases can be accessed in a number
of different ways: through the ESO science archive, from the
{\sc Astro-WISE} archive, or via a web-based synoptic table on the KiDS
website\footnote{\url{http://kids.strw.leidenuniv.nl/DR2}}.

\subsubsection{ESO archive}

As an ESO public survey, the KiDS data releases are distributed via
the ESO Science Archive
Facility\footnote{\url{http://archive.eso.org}}. Using the Phase 3
query forms users can find and download data products such as the
stacked images and source lists.

The naming convention used for all data product files in the ESO archive is the following: \\
{\tt KiDS\_DRV.V\_R.R\_D.D\_F\_TTT.fits},\\
where {\tt V.V} is the data release version, {\tt R.R} and {\tt D.D}
are the RA and DEC of the tile center in degrees (J2000.0) with 1
decimal place, {\tt F} is the filter ($u$, $g$, $r$, $i$, or $ugri$),
and {\tt TTT} is the data product type. Table \ref{Tab:FileTypes}
lists the ESO product category name, file type, value of {\tt TTT}, 
and an example filename for each type of data product.

For example, the KiDS-ESO-DR1 $r$-band stacked image of the tile
"KIDS\_48.3\_-33.1" is called {\tt
  KiDS\_DR1.0\_48.3\_-33.1\_r\_sci.fits}, and the KiDS-ESO-DR2
multi-band source list data file corresponding to this tile is called {\tt
  KiDS\_DR2.0\_48.3\_-33.1\_ugri\_src.fits}.

\begin{table*}
\centering
\caption{Data products and file types.}
\label{Tab:FileTypes}
\begin{tabular}{l l l l}
\hline\hline
Data product & ESO product category name & File type & TTT\tablefootmark{a} \\
\hline
Calibrated, stacked images & SCIENCE.IMAGE & FITS image & sci \\
Weight maps & ANCILLARY.WEIGHTMAP & FITS image & wei \\
Masks & ANCILLARY.MASK & FITS image & msk \\
Single-band source lists & SCIENCE.SRCTBL & Binary FITS table & src \\
Multi-band catalogue & SCIENCE.SRCTBL & Binary FITS table & src \\
\hline\\
\end{tabular}
\tablefoot{
\tablefoottext{a}{TTT is the three character string indicating the data product type.}
}
\end{table*}

\subsubsection{Astro-WISE archive}

All data products can also be retrieved from the {\sc Astro-WISE} system \citep{awsystem,awexpast},
the main data processing and management system used for processing
KiDS data. The pixel data and source lists are identical to the data
stored in the ESO Science Archive Facility, but additionally, the full
data lineage is available in {\sc Astro-WISE}. This makes this access route convenient for those wanting to access the various quality controls or further analyse, process or data-mine the full data set instead of particular tiles of interest.

All data products can be accessed through the links provided on the
KiDS website via the DBviewer interface. Downloading of files, viewing
inspection plots, and browsing the data lineage is fully supported by
the web interface.

\subsubsection{Synoptic table}

A third gateway to the KiDS data is the synoptic table that is
included on the KiDS website. In this table quality information on
different data products is combined, often in the form of inspection
figures, offering a broad overview of data quality. 

\section{Observational set-up}
\label{Sec:ObservationalSetup}

The KiDS survey area is split into two fields, KiDS-North and KiDS-South, covering a large range in right ascension so that observations can be made all year round. The fields, each approximately 750 square degrees in size, were chosen to overlap with several large galaxy redshift surveys, principally SDSS, the 2dF Galaxy Redshift Survey \citep[2dFGRS, ][]{colless+01} and the Galaxy And Mass Assembly (GAMA) survey \citep{gama}. KiDS-North is completely covered by the combination of SDSS and the 2dFGRS, while KiDS-South corresponds to the 2dFGRS south Galactic cap region. Four out of five GAMA fields lie within the KiDS fields. Figure \ref{Fig:SkyDistribution} shows the outline of the survey fields.

Each survey tile is observed in the $u$, $g$, $r$, and $i$ bands. Exposure
times and observing constraints for the four filters are designed to
match the atmospheric conditions on Paranal and optimized for the
survey's main scientific goal of weak gravitational lensing. 
KiDS makes use of queue scheduling, allowing the data requiring the best conditions to be observed whenever these conditions are met. In order to promote building up full wavelength coverage as quickly as possible, pointings for which a subset of filters has been observed receive higher priority.  Unfortunately, the queue scheduling system does not allow prioritizing of survey tiles based on the observational progress of neighboring tiles. As a result, the queued tiles are observed in a random order, resulting in the patchy on-sky distribution visible in Fig.~\ref{Fig:SkyDistribution}.
The median galaxy redshift of the final survey will reach 0.7 and the best
seeing conditions are reserved for the $r$-band, since this
functions as the shape measurement band. Each position on the sky is
visited only once in each filter, so that the full survey depth is
reached immediately. While this precludes variability studies, it will
allow the other science projects to benefit from deep data from the
start. During the survey the observing constraints have been fine-tuned to optimize between survey speed and scientific return, and Table \ref{Tab:ObservingConstraints} lists the limits
employed for the majority of the period during which the data
presented here were obtained.

\begin{figure}
   \centering
   \includegraphics[width=8cm]{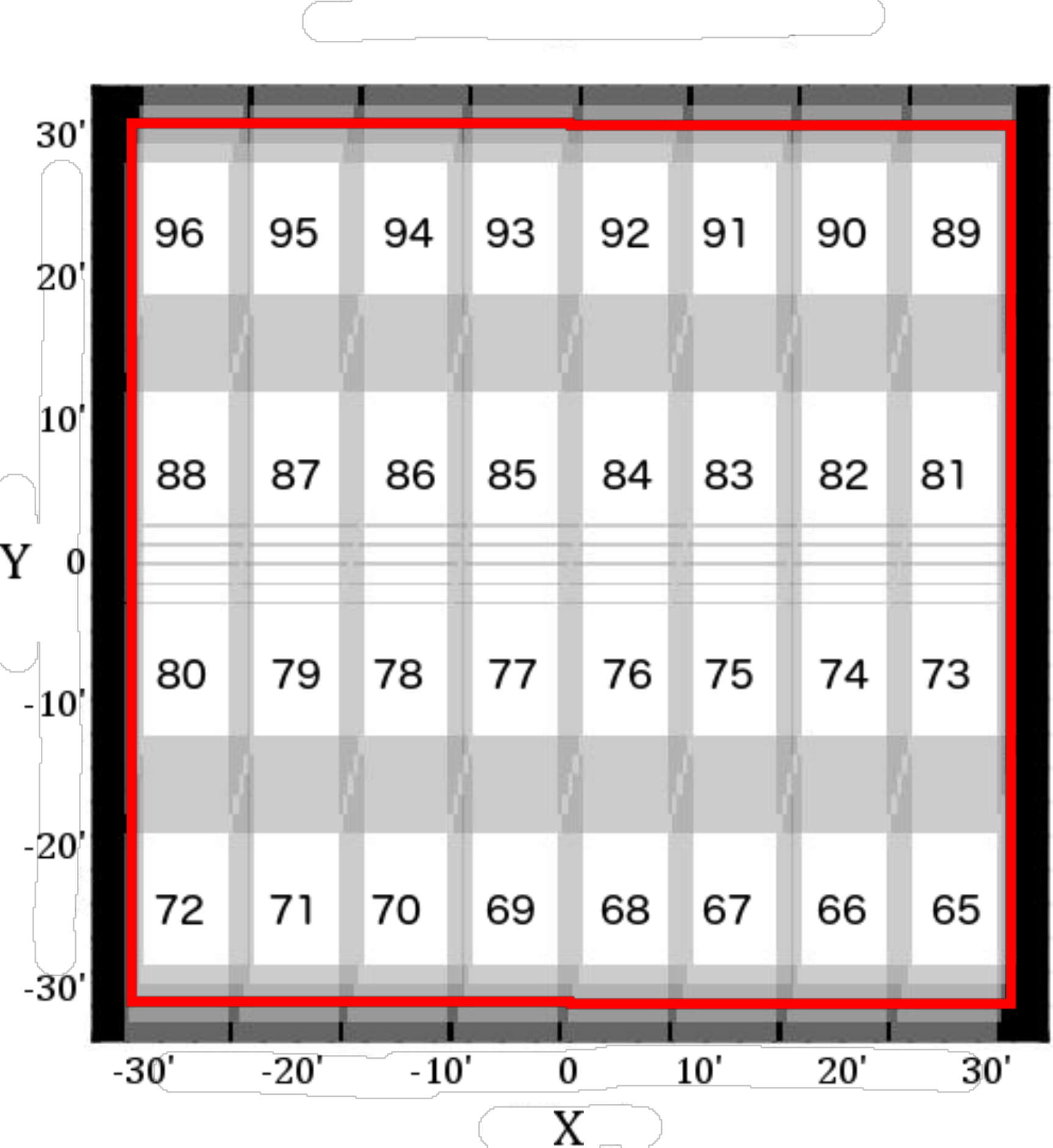}
   \caption{Tile footprint sky coverage for observations in $g$, $r$ and $i$. To fill in the gaps between the CCDs five dither steps are done, yielding a sky coverage pattern where the vast majority of the field-of-view is covered by four or five exposures. White regions are covered by five exposures and increasingly darker shades of gray by four, three, two and one. The numbers 65 to 96 indicate the CCD numbering according to the CCD naming scheme and the red rectangle shows how tiles are cropped while creating the multi-band catalogue.}
    \label{Fig:DithersFootprint}
    \end{figure}

Since the OmegaCAM CCD mosaic consists of 32 individual CCDs, the sky
covered by a single exposure is not contiguous but contains gaps. In
order to fill in these gaps, KiDS tiles are built up from 5 dithered
observations in $g$, $r$ and $i$ and 4 in $u$. The dithers form a
staircase pattern with dither steps of 25\arcsec\ in X (RA) and
85\arcsec\ in Y (DEC), bridging the inter-CCD gaps
\citep{kids_expast}, see Fig. \ref{Fig:DithersFootprint}. Although
filled in, the gaps result in areas within the footprint that are
covered by fewer than 5 exposures, thus yielding slightly lower
sensitivity. Due to this dithering strategy the final footprint of
each tile is slightly larger than 1 square degree: 
$61.9\times65.4$ arcminutes in $u$; $62.3\times66.8$ arcminutes in $g$, $r$
and $i$.  Neighbouring dithered stacks have an overlap in RA of 5\%
and in DEC of 10\%. The tile centers are based on a tiling strategy
that covers the full sky efficiently for VST/OmegaCAM
\footnote{\url{http://www.astro.rug.nl/~omegacam/dataReduction/Tilingpaper.html}}.
The combination of the tiling and dithering scheme ensures that every point within the survey area is covered by a minimum of 3 exposures.

\begin{table*}
\caption{KiDS observing strategy: observing condition constraints and exposure times.}
\label{Tab:ObservingConstraints}
\centering
\begin{tabular}{l c c c c c c c}
\hline\hline
Filter & Max. lunar & Min. moon & Max. seeing & Max. airmass & Sky transp. & Dithers & Total Exp.\\
~ & illumination & distance (deg) & (arcsec) & ~ & ~ & ~ & time (s) \\
\hline
$u$ & 0.4 & 90 & 1.1 & 1.2 & CLEAR & 4 & 1000 \\
$g$ & 0.4 & 80 & 0.9 & 1.6 & CLEAR & 5 & 900 \\
$r$ & 0.4 & 60 & 0.8 & 1.3 & CLEAR & 5 & 1800 \\
$i$ & 1.0 & 60 & 1.1 & 2.0 & CLEAR & 5 & 1200 \\
\hline
\end{tabular}
\end{table*}

\section{Data processing}
\label{Sec:DataProcessing}

The data processing pipeline used for KiDS-ESO-DR1/2 is
based on the {\sc Astro-WISE} optical pipeline described in
\citet[henceforth MF13]{astrowise}. Below we summarize the processing steps and
list KiDS-specific information, covering the KiDS process configuration and
departures from the {\sc Astro-WISE} optical pipeline.

\subsection{Image detrending}
\label{Sec:ImageDetrending}

The first processing steps are the detrending of the raw data, consisting of the following steps.\\

\begin{figure*}
\centering
\includegraphics[width=\textwidth]{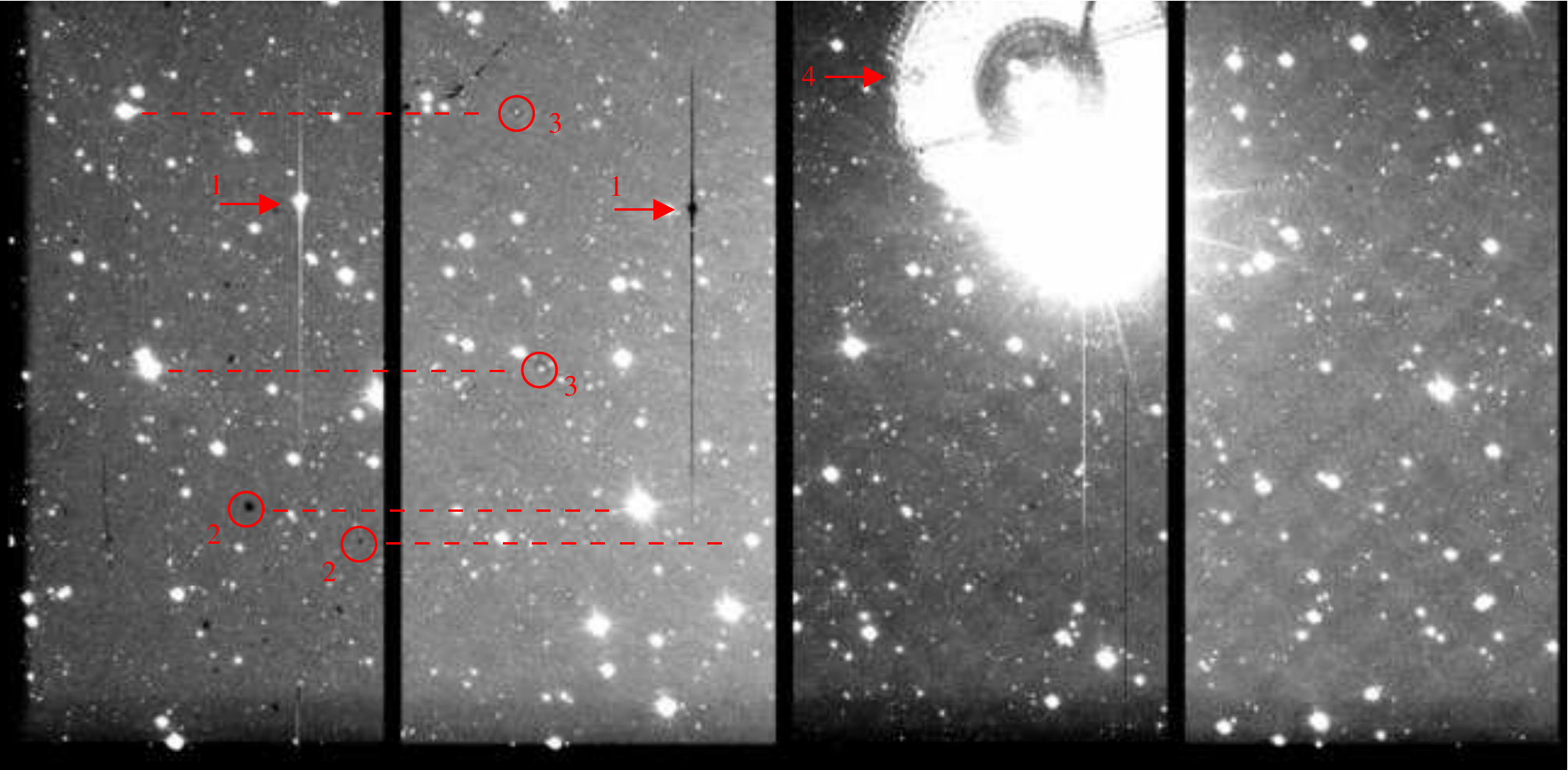}
\caption{Example of crosstalk between CCDs 96, 95, 94 and 93 (left to right) of OmegaCAM. Both positive and negative crosstalk due to the bright star in CCD 94 is visible (1). Negative crosstalk on CCD 96 due to bright (saturated and unsaturated) stars on CCD 95 is indicated (2), as well as positive crosstalk on CCD 95 due to bright stars on CCD 96 (3). The bright ring on CCD 94 (4) is an optical "ghost" reflection.}
\label{Fig:CrosstalkExample}
\end{figure*}

\noindent {\bf Cross-talk correction.} Electronic cross-talk occurs
between several CCDs, but most strongly between CCDs \#93, \#94, \#95
and \#96, which share the same video board (see Fig.~\ref{Fig:DithersFootprint} 
for CCD numbering scheme). Cross-talk can be both
positive and negative, resulting in faint imprints of bright sources
on neighbouring CCDs (Fig.~\ref{Fig:CrosstalkExample}).
Although the cross-talk is generally stable,
abrupt changes can occur during maintenance or changes to the
instrumentation.

\begin{figure}
\includegraphics[width=\columnwidth]{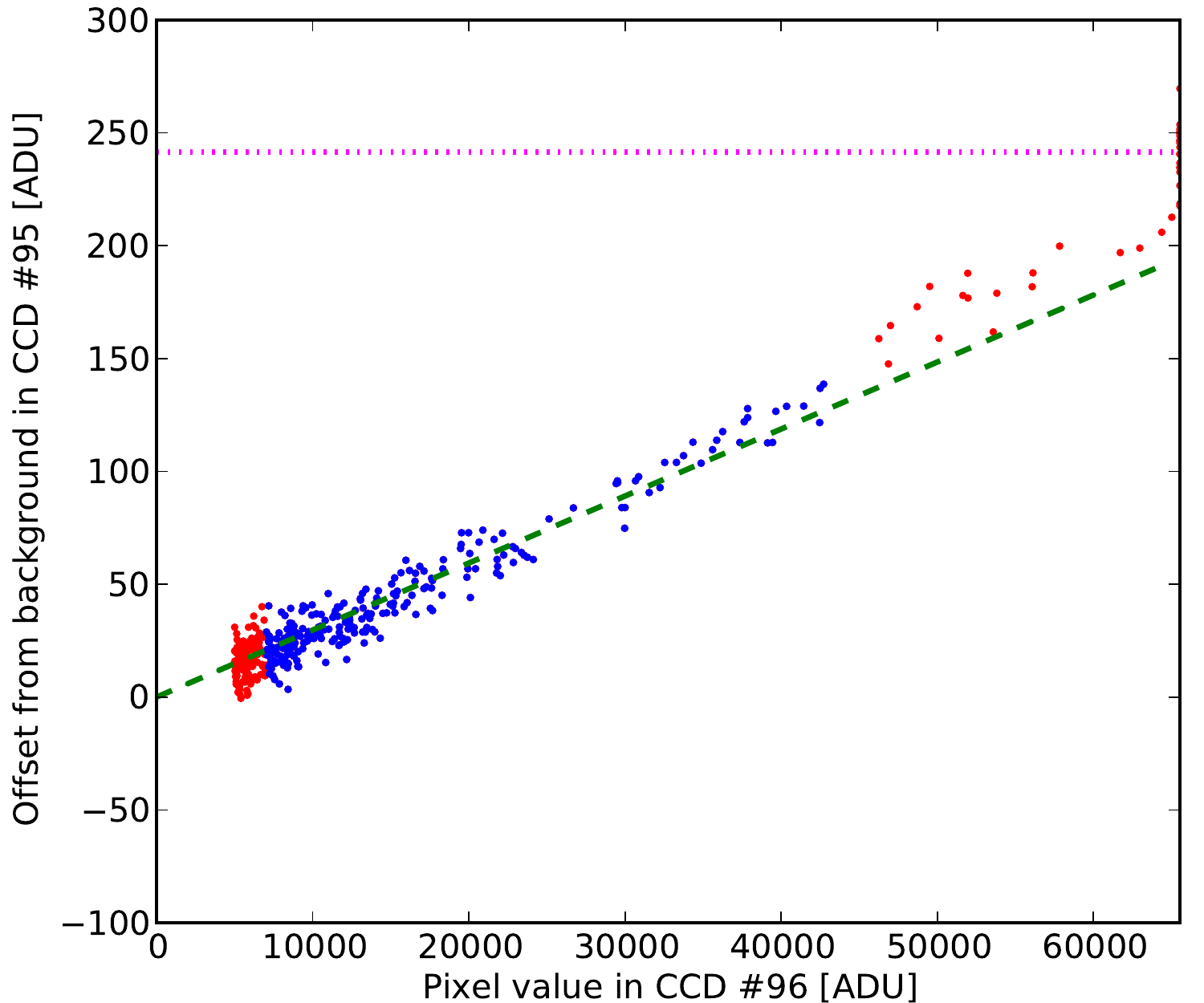}
\caption{Measurement of crosstalk between CCDs \#95 and \#96 for one exposure. A straight line (green dashed) is fit to the offset from the median in the target CCD (here: \#95) as a function of signal in the same pixel in the source CCD (here: \#96). Only the blue data points are used for this fit since at low ADU levels outliers are often present and at high ADU levels non-linearity affects the measurements. The slope of the green dashed line corresponds to the correction coefficient $b$. The magenta dotted line shows the constant $a$ value used for saturated pixels.}
\label{Fig:CrosstalkMeasurement}
\end{figure}

A correction is made for cross-talk between CCDs \#95 and \#96, where
it is strongest (up to 0.7\%). Crosstalk between a pair (``source''
and ``target'') of CCDs is determined by measuring for each pixel, with a
value greater than 5000 ADU in the source CCD, the offset of the same
pixel in the target CCD from the median value of all pixels in the
target CCD. A straight line is fit to the trend of this offset
as function of the pixel value of the source CCD. See Fig.
\ref{Fig:CrosstalkMeasurement} for an example. The slope of the line
($b$) is given in Table \ref{Tab:CrosstalkCoefficients} per stable
period.  A separate constant is fit to saturated pixels in the
source CCD ($a$ in the table).  To correct for the crosstalk between
CCDs \#95 and \#96 the correction factor is applied to each pixel in
the target CCD based on the corresponding pixel values in the source
CCD:
\begin{equation}
I'_i = 
\begin{cases}
I_i + a, &\text{if $I_j = I_{\rm{sat.}}$;}\\
I_i + b  I_j, &\text{if $I_j < I_{\rm{sat.}}$,}\\
\end{cases}
\end{equation}
where $I_i$ and $I_j$ are the pixel values in CCDs $i$ and $j$, $I'_i$ is the corrected pixel value in CCD $i$ due to cross-talk from CCD $j$, and $I_{\rm{sat.}}$ is the saturation pixel value.

\begin{table}
\caption{Applied cross-talk coefficients.}
\label{Tab:CrosstalkCoefficients}
\centering
\footnotesize
\resizebox{\columnwidth}{!}{
\begin{tabular}{l | c c | c c}
\hline\hline
Period & \multicolumn{2}{c|}{95 to 96} & \multicolumn{2}{c}{96 to 95} \\
~ & $a$ & $b$ ($\times10^{-3}$) & $a$ & $b$ ($\times10^{-3}$)\\
\hline
2011-08-01 - 2011-09-17 & $-$210.1 & $-$2.504 & 59.44 & 0.274 \\
2011-09-17 - 2011-12-23 & $-$413.1 & $-$6.879 & 234.8 & 2.728 \\
2011-12-23 - 2012-01-05 & $-$268.0 & $-$5.153 & 154.3 & 1.225 \\
2012-01-05 - 2012-07-14 & $-$499.9 & $-$7.836 & 248.9 & 3.110 \\
2012-07-14 - 2012-11-24 & $-$450.9 & $-$6.932 & 220.7 & 2.534 \\
2012-11-24 - 2013-01-09 & $-$493.1 & $-$7.231 & 230.3 & 2.722 \\
2013-01-09 - 2013-01-31 & $-$554.2 & $-$7.520 & 211.9 & 2.609 \\
2013-01-31 - 2013-05-10 & $-$483.7 & $-$7.074 & 224.7 & 2.628 \\
2013-05-10 - 2013-06-24 & $-$479.1 & $-$6.979 & 221.1 & 2.638 \\
2013-06-24 - 2013-07-14 & $-$570.0 & $-$7.711 & 228.9 & 2.839 \\
2013-07-14 - 2014-01-01 & $-$535.6 & $-$7.498 & 218.9 & 2.701 \\
\hline
\end{tabular}
}
\normalsize
\end{table}

\noindent {\bf De-biasing and overscan correction.}
The detector bias is subtracted from the KiDS data in a two step
procedure. First, for each science and calibration exposure the
overscan is subtracted per row (no binning of rows). For consistency, all science and calibration data are reduced with this same overscan correction method. Second, a
daily overscan-subtracted master bias, constructed from ten bias frames and applying 3$\sigma$ rejection, is subtracted.

\begin{figure*}
\centering
\includegraphics[width=\textwidth]{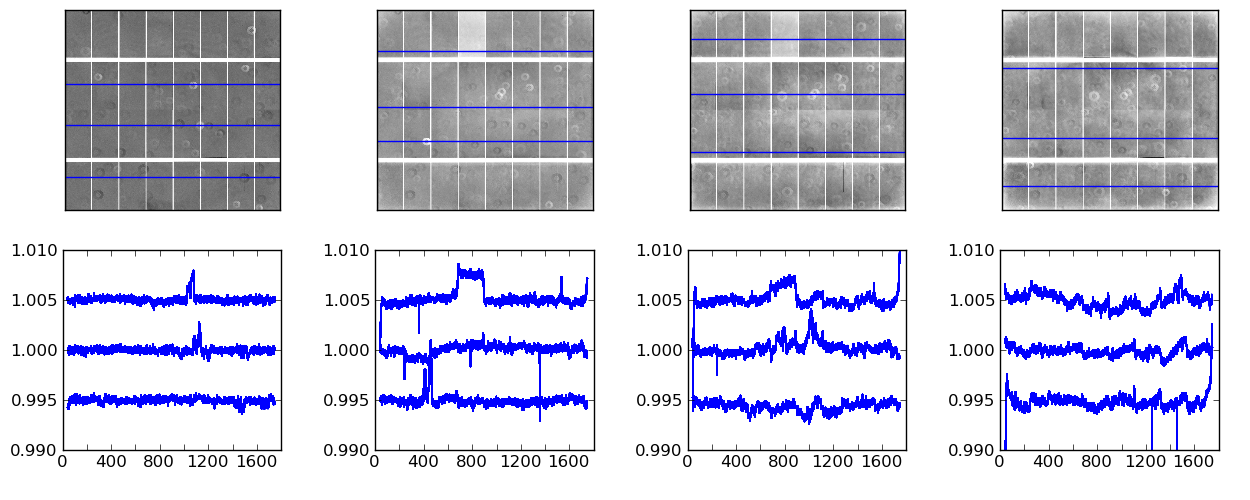}
\caption{Dome flat variation over time during the observing period of KiDS-ESO-DR2. The top panels show the $10\times10$ binned dome flat ratios between a dome flat taken on August 7 2011 and, from left to right, October 1 2011, April 1 2012, March 2 2013, and November 2 2013. The bottom panels show three cuts through the ratio images on top, along the lines indicated in blue on the corresponding top panels. Offsets of $-0.005$ and $+0.005$ are applied to the bottom and top lines, respectively, to improve legibility.}
\label{Fig:DomeflatVariation}
\end{figure*}

\noindent {\bf Flat-fielding.}
A single masterflat (per CCD and filter) was used for all data in the
release. This is by virtue that the intrinsic pixel sensitivities can
be considered constant to $\sim0.2\%$ or better for $g$, $r$ and $i$
\citep{verdoes13}. 
This is illustrated in Fig. \ref{Fig:DomeflatVariation}, where a series of dome flat ratios is shown. The dome flatfields offer the optimal controlled experiment for assessing the pixel sensitivities, as the conditions under which they are observed are closely monitored and calibrated. More specifically, 
the calibration unit has a special power supply which allows the ramping of the current when switching on and off and the stabilization of the current during the exposure, delivering an exposure level variation less than 0.6 \% over a period of 1 month \citep{verdoes13}.
The series of ratio plots in Fig. \ref{Fig:DomeflatVariation} spans the full time period during the KiDS-ESO-DR1/2 observations, demonstrating that the peak-to-valley pixel-to-pixel variations vary less than 0.5\% at any time and pixel position.\\
For $g$, $r$ and $i$ the master flat is a combination of a master
dome (for high spatial frequencies) and master twilight (=sky)
flat-field (for low spatial frequencies). Both contributing flats are
an average of 5 raw flat-field exposures with 3$\sigma$ rejection. In
$u$ band only the twilight flats are used.

\noindent{\bf Illumination correction.} 
An illumination correction (also known as ``photometric superflat'') is required to correct for illumination variations due to stray light in the flat field images, because of which the flat fields are not a correct representation of the pixel sensitivities and vignetting effects. The illumination correction is applied
{\it in pixel space}, and only on the source fluxes (i.e., after
background subtraction). A single illumination correction image is
used to correct the single master flats per filter for all data. The correction is determined from observations of several Landolt Selected Area (SA) fields \citep{landolt92} observed at 33 dither positions, such that the same stars are observed with all CCDs. After computing zero-points per CCD the residuals between the measured stellar magnitudes and their reference values sample the illumination variations. Per CCD the illumination correction is characterized to better than 1\%. 
For further details see \cite{verdoes13}.

\noindent {\bf De-fringing.} 
De-fringing is only needed for the KiDS $i$-band. Analysis of nightly
fringe frames showed that the pattern is constant in time. Therefore,
a single fringe image was used for all KiDS-ESO-DR1 images observed
after 2012-01-11. For each science exposure this fringe image is
scaled and then subtracted to minimize residual fringes.
The general procedure, described in MF13, was modified for the KiDS
data in order to take large-scale background fluctuations into
account. Some $i$-band data contain significant background fluctuations
even after flatfielding due to scattered light (see Sect. 
\ref{Sec:DataFoibles}), which can cause problems with the scaling of the
fringe frame. Therefore, background-subtracted science frames were
used to determine the scale factor.

\noindent {\bf Pixel masking.} 
Cosmic-rays, hot and cold pixels, and saturated pixels are automatically
masked as described in MF13 during de-trending. These are included in
the weight image. Additional automatic and manual masking is then applied
on the coadds as described in Sect.~\ref{Sec:Masking}.

\noindent {\bf Satellite track removal.} 
Satellite tracks are detected by an automated procedure, working on
single CCDs, that applies the Hough transform \citep{hough62} to a
difference image between exposures with the smallest dither offset. 
Bright stars and bright reflection halos are
masked to limit false detections. The pixels affected by satellite
tracks are masked and included in the weight image. Due to the
single-CCD approach small sections of satellite tracks in CCD corners
may be missed by the algorithm. These remnants are included in the
manual masks discussed in Sect.~\ref{Sec:Masking}.

\noindent {\bf Background subtraction.}  
Many observations show darkened regions at the horizontal CCD edges
where the bond wire baffling is placed (Iwert et al. 2006). For 
KiDS-ESO-DR1/2 this CCD-edge vignetting is corrected by
performing a line-by-line background subtraction, implemented as a new
background subtraction method during regridding (see Sect. 3.5 of
MF13).

The line-by-line background subtraction can be inaccurate if there are
not enough `background' pixels on a line, for example nearby bright
stars. These regions are identified in the masks.

\noindent {\bf Non-linearity.}
The linearity of the response of the OmegaCAM detector-amplifier chain is regularly monitored as part of the VST calibration procedures. No significant non-linearities are present (at the level of $\lesssim$1\% over the full dynamic range of the system), and the pipeline currently does not include a non-linearity correction step.

\subsection{Photometric calibration}
\label{Sec:PhotometricCalibration}

The steps taken to calibrate the photometry are as follows. The calibration described here is performed per tile and per filter and applied to the pixel data, which is rescaled to the flux scale in Eq.~\ref{Eq:fluxscale}.

\begin{itemize}

\item Photometric calibration of the KiDS-ESO-DR1/2
  data starts with individual zero-points per CCD based on SA field
  observations (see MF13 for details of the zero-point
  derivation). The calibration deploys a fixed aperture of 30 pixels ($\sim$6.4\arcsec\
  diameter) not corrected for flux losses, and uses SDSS DR8 PSF
  magnitudes of stars in the SA fields as reference. 
  Zero-points are determined every night, except when no SA field observations are available, in which case default values, that were determined in the same way, are used. The total number of tiles for which these default values were used is 16, 13 and 14 in $g$, $r$ and $i$, respectively, or $\sim$10\% of all tiles. For $u$ the nightly zero-point determinations often show large uncertainties and default zero-points are used more frequently, namely in 112 tiles.
  Magnitudes are
  expressed in AB in the instrumental photometric system. No
  colour corrections between the OmegaCAM and SDSS photometric system
  are applied. 

\item Next, the photometry in the $g$, $r$ and $i$ filters is
  homogenized across CCDs and dithers for each tile independently per filter. For
  $u$-band this homogenization is not applied because the relatively
  small source density often provides insufficient information to tie
  adjacent CCDs together. This global photometric solution is derived
  and applied in three steps:

\begin{enumerate}
\item From the overlapping sources across dithered exposures, photometric
  differences between the dithers (e.g. due to varying
  atmospheric extinction) are derived.
\item Although each CCD was calibrated with its own zero-point, the 
  remaining photometric differences  between CCDs are calculated 
  using all CCD overlaps between the dithered exposures. The number of sources in these overlaps can range between a few to several hundreds, depending on the filter and the size of the overlap, and a weighting scheme is used based on the number of sources. Both in 
  steps 1 and 2, photometric offsets are obtained by minimizing the 
  difference in source fluxes between exposures and CCDs using the 
  algorithm of  \cite{maddox90}.
\item The offsets are applied to all CCDs with respect to
  the zero-points valid for the night, derived from the nightly
  SA field observations.
\end{enumerate}

\end{itemize}

\subsection{Astrometric calibration}
\label{Sec:AstrometricCalibration}

A global (multi-CCD and multi-dither) astrometric calibration is
calculated per filter per tile. {\sc SCAMP} \citep{scamp} is used for
this purpose, with a polynomial degree of 2 over the whole mosaic. The
(unfiltered) 2MASS-PSC \citep{skrutskie06} is used as astrometric
reference catalogue, thus the astrometric reference frame used is International Celestial Reference System (ICRS). Using unwindowed positions the external (i.e. with
respect to the 2MASS-PSC) and internal accuracies are typically
described by a 2-D RMS of 0.3\arcsec and 0.03\arcsec, respectively
(see also Sect. \ref{Sec:AstrometricQuality}). A more detailed description
of the astrometric pipeline can be found in MF13.

\subsubsection{Regridding and coadding}

{\sc SWARP} \citep{swarp} is used to resample all exposures in a tile
to the same pixel grid, with a pixel size of 0.2\arcsec. After an
additional background subtraction step (using 3$\times$3 filtered
128$\times$128 pixel blocks) the exposures are scaled to an effective
zero-point of 0 (see also Sect. \ref{Sec:coaddunits}) and coadded using a
weighted mean stacking procedure. The applied projection method is Tangential, Conic-Equal-Area. Due to the individual photometric offsets applied to the CCDs the gain varies slightly over the
coadd. An average effective gain is calculated for each coadd as:
\begin{equation}
<G> = N_{\rm{exp}} \times \frac{\tilde{G}}{\tilde{S}},
\end{equation}
where $\tilde{G}$ and $\tilde{S}$ are the median gain and
scale factor of the regridded CCD images and $N_{\rm{exp}}$ is the number of
exposures. These average gain values are provided in the 
online Data table\footnote{\url{http://kids.strw.leidenuniv.nl/DR2}}.

\subsection{Masking of bright stars and defects}
\label{Sec:Masking}

   \begin{figure*}
   \centering
   \includegraphics[width=18cm]{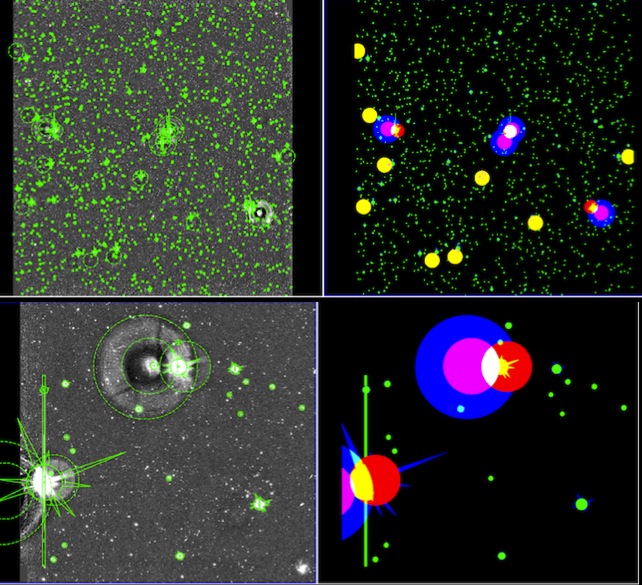}
   \caption{Examples of a {\sc Pulecenella} v1.0 mask. {\em Top left}: thumbnail of a KiDS stacked image with several bright stars; critical areas detected by {\sc Pulecenella} are overplotted with green outlines (region format). {\em Top right}: FLAG image corresponding to the image shown on the top left, with different colours corresponding to the pixel value with the binary flag values summed. {\em Bottom} panels: as top panels, but for a small area in a (different) KiDS stacked image.}
    \label{Fig:MaskExample}
    \end{figure*}

Nearby saturated stars and other image defects are the main source of
contamination in the measurement of objects. In KiDS coadds, these
features are masked by a stand-alone program, {\sc Pulecenella}
(v1.0, Huang et al. in prep). {\sc Pulecenella} is a novel procedure
for automated mask creation completely independent of external star
catalogues. An example of a mask image is shown in Fig.
\ref{Fig:MaskExample}.

{\sc Pulecenella} detects and classifies the following types of image
artifacts resulting from the saturated stars:
\begin{itemize}
\item[--] saturated pixels in the core and vicinity of stars,
\item[--] spikes caused by diffraction of the mirror supports,
\item[--] spikes caused by readout of saturated pixels,
\item[--] ``ghost'' halos produced by the reflections off of optics (up to three wider reflection halos with spatially dependent offsets; these are caused by reflections of different optical elements in the light-path and depend on the brightness of the star).
\end{itemize}

These features have regular shapes, and scale with the brightness and
position of bright stars in a stable way in all images in a given
observation band. Hence, {\sc Pulecenella} is first configured to model the
mask shapes (including the radius of saturation cores, the orientation
of diffraction spikes, the size and offset of reflection halos) from
some sampled saturated stars; the so configured analytical models are
then applied to the batch masking of coadds from the same band. The
detection, location and magnitude of saturated stars are derived from
a first {\sc SExtractor} \citep{sextractor} run over the image to be masked, with a
band-specific configuration aiming only for the detection and
measurement of the nearly saturated stars; the saturation pixel level
is derived from the FITS image header. Thus, {\sc Pulecenella} produces
star masks specifically for each image, without any dependence on an
external star catalogue. This also avoids the ambiguity in the
determination of the magnitude cutoff of saturated stars due to the
difference between the observed and external catalogue filters.

The position of reflection halos is offset from the center of the host
saturated star; the offset can be towards or outward from the image
center, depending on the reflection components in the optics. For the
primary reflection halo, the offset is first linearly modeled from
several primary halos of the brightest stars in the image, and whether
the halo mask is applied depends on the brightness of the saturated
stars; this brightness level is determined as the number of pixels for
which the count level exceeds 80\% of the saturation limit and is
therefore independent of the photometric calibration. The same method is applied
for secondary and tertiary halos independently. The details of the
masking method are presented in Huang et al. (2015,
in prep).

Besides the analytical star masks, {\sc Pulecenella} is also configured to
mask pixels in the empty boundary margins, CCD gaps or dead pixels
with zero weight. These pixels are flagged as 'bad pixels'.  
Finally, bad regions which are missed by {\sc Pulecenella}
(e.g. large-scale background artifacts) but are detected by a visual
inspection, are manually masked using DS9 and added both to the region
file and to the final flag image.

As output, {\sc Pulecenella} generates both an ASCII mask region file which
is compatible with the {\sc DS9}\footnote{\url{http://ds9.si.edu}} tool and a FITS flag image that can be used
in {\sc SExtractor}. Different types of masked artifacts/regions are coded
with the different binary values listed in Table
\ref{Tab:MaskValues}. In the flag image these binary values are
summed.

The flag image is used during source extraction for the single-band
source list (see below) to flag sources whose isophotes overlap with
the critical areas. The resulting flags are stored in the following
two {\sc SExtractor} parameters:

\begin{itemize}
\item[--] IMAFLAGS$\_$ISO: sum of all mask flags encountered in the isophote profile,
\item[--] NIMAFLAG$\_$ISO: number of flagged pixels entering IMAFLAGS$\_$ISO.
\end{itemize}

   \begin{figure*}[ht]
   \centering
   \includegraphics[width=8.5cm]{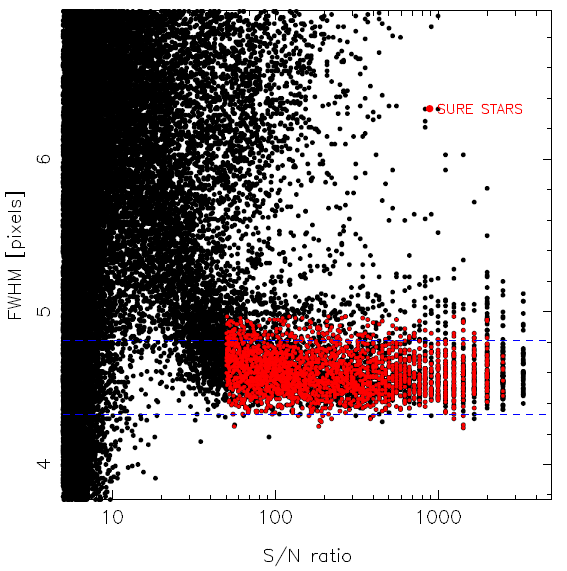}
   \includegraphics[width=8.5cm]{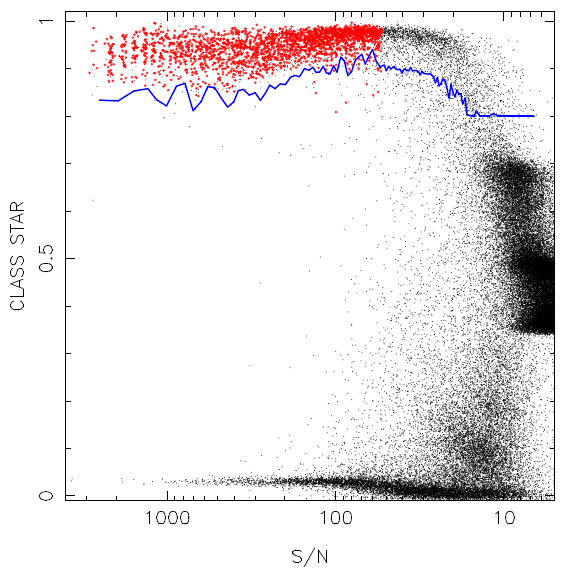}
   \caption{High-confidence star candidates and star/galaxy separation. {\it Left:} the high-confidence
star candidates (red dots) are used to locate the stellar locus and calculate the average FWHM of
the image. {\it Right:} example of star/galaxy separation; at SNR >50, the high-confidence star candidates
(red dots) are used to define the blue line; at lower SNR, all sources with CLASS\_STAR>0.8 are
used; sources above the blue line are classified as stars.}
    \label{Fig:SGseparation}
    \end{figure*}

\begin{table}
\caption{Critical areas in the masks and their flag values.}
\label{Tab:MaskValues}
\centering
\begin{tabular}{l c | l c}
\hline\hline
Type of area & Flag & Type of area & Flag \\
\hline
Readout spike & 1 & Secondary halo & 16 \\
Saturation core & 2 & Tertiary halo & 32 \\
Diffraction spike & 4 & Bad pixel & 64 \\
Primary halo & 8 & Manually masked & 128 \\
\hline
\end{tabular}
\end{table}

Table \ref{Tab:AreaMasked} summarizes the percentages of the total area in KiDS-ESO-DR1/2 that are not masked, masked automatically by {\sc Pulecenella}, and manually masked. Due to the lower sensitivity in $u$ the number of saturated stars is much smaller, leading to a significantly smaller percentage of masked pixels. The high fraction of area that is manually masked in $i$-band is due to the higher frequency and severity of scattered light issues (see Sect. \ref{Sec:DataFoibles}) caused by moonlight and higher sky brightness. Taking into account the nominal area covered by KiDS-ESO-DR1/2 the total unmasked area is approximately 120 square degrees.

\begin{table}
\caption{Percentage of masked area.}
\label{Tab:AreaMasked}
\centering
\begin{tabular}{l c c c}
\hline\hline
Filter & Not masked & Automatically & Manually \\
~ & ~ & masked & masked \\
\hline
$u$ & 97 & 1 & 2 \\
$g$ & 86 & 7 & 7 \\
$r$ & 78 & 12 & 10 \\
$i$ & 77 & 9 & 14 \\
\hline
\end{tabular}
\end{table}

\subsection{Source extraction and star/galaxy separation}
\label{Sec:SourceExtraction}

Single-band source lists were included in KiDS-ESO-DR1, while KiDS-ESO-DR2 contains both single-band source lists, as well as a multi-band source catalogue encompassing the combined area. The single-band source lists are provided per survey tile, while the catalogue, although split into files corresponding to single tiles, is constructed as a single catalogue. Both the source lists and the catalogue are intended as ``general purpose'' catalogues.

\subsubsection{Single-band source lists}
\label{Sec:SingleBandSourceLists}

\begin{table*}
\centering
\caption{Detection set-up for KiDS-ESO-DR1 and KiDS-ESO-DR2 single-band source lists.}
\label{Tab:SExtractorSetup}
\begin{tabular}{l c l}
\hline\hline
Parameter & Value & Description \\
\hline
{\tt DETECT\_THRESH} & 1.5 & <sigmas> or <threshold>,<ZP> in mag/arcsec$^2$ \\
{\tt DETECT\_MINAREA} & 3 & minimum number of pixels above threshold \\
{\tt ANALYSIS\_THRESH} & 1.5 & <sigmas> or <threshold>,<ZP> in mag/arcsec$^2$ \\
{\tt DEBLEND\_NTHRESH} & 32 & Number of deblending sub-thresholds \\
{\tt DEBLEND\_MINCONT} & 0.001 & Minimum contrast parameter for deblending \\
{\tt FILTER} & Y & Apply filter for detection (Y or N) \\
{\tt FILTER\_NAME} & default.conv & Name of the file containing the filter \\
{\tt CLEAN} & Y & Clean spurious detections? (Y or N) \\
{\tt CLEAN\_PARAM} & 1.0 & Cleaning efficiency \\
{\tt BACK\_SIZE} & 256 & Background mesh: <size> or <width>,<height> \\
{\tt BACK\_FILTERSIZE} & 3 & Background filter: <size> or <width>,<height> \\
{\tt BACKPHOTO\_TYPE} & LOCAL & can be GLOBAL or LOCAL \\
{\tt BACKPHOTO\_THICK} & 24 & thickness of the background LOCAL annulus \\
\hline
\end{tabular}
\end{table*}

Source list extraction and star/galaxy (hereafter S/G) separation is achieved with an automated
stand-alone procedure optimized for KiDS data: {\sc KiDS-CAT}. This procedure, the backbone of which is formed by {\sc SExtractor}, performs the following
steps separately for each filter.
\begin{enumerate}
\item {\sc SExtractor} is run on the stacked image to measure the FWHM of all sources. High-confidence
star candidates are then identified based on a number of criteria including signal-to-noise ratio (SNR) and ellipticity cuts \citep[for details see][]{labarbera08}.
\item The average PSF FWHM is calculated by applying the bi-weight location estimator to the
FWHM distribution of the high-confidence star candidates.
\item A second pass of {\sc SExtractor} is run with SEEING\_FWHM set to the derived average PSF
FWHM. During this second pass the image is background-subtracted, filtered and thresholded
``on the fly''. Detected sources are then de-blended, cleaned, photometered, and
classified. A number of {\sc SExtractor} input parameters are set individually for each image
(e.g., SEEING\_FWHM and GAIN), while others have been optimized to provide the best
compromise between completeness and spurious detections (see Data Quality section
below). The detection set-up used is summarized in Table \ref{Tab:SExtractorSetup}; a full {\sc SExtractor} configuration file
is available online\footnote{\url{http://kids.strw.leidenuniv.nl/DR1/example\_config.sex}}.
Apart from isophotal magnitudes and Kron-like elliptical aperture magnitudes, a large
number of aperture fluxes are included in the source lists. This allows users to estimate
aperture corrections and total source magnitudes. All parameters provided in the source
lists are listed in the Data Format section below.
\item S/G separation is performed based on the CLASS\_STAR (star classification) and SNR (signal-to-noise ratio) parameters provided by {\sc SExtractor} and consists of the following steps:
\begin{itemize}
\item In the SNR range where the high-confidence star candidates are located (the red
dots in Fig. \ref{Fig:SGseparation}) the bi-weight estimator is used to define their CLASS\_STAR location,
$\theta$, and its width, $\sigma$; a lower envelope of $\theta - 4\sigma$ is defined.
\item At SNR below that of the high-confidence star candidates, a running median
CLASS\_STAR value is computed based on all sources with CLASS\_STAR > 0.8. This running median is
shifted downwards to match the $\theta - 4\sigma$ locus. The resulting curve (blue curve in Fig. \ref{Fig:SGseparation})
defines the separation of stars and galaxies.
\end{itemize}
\end{enumerate}

The source magnitudes and fluxes in the final source lists are for zero airmass, but not
corrected for Galactic foreground or intergalactic extinction. The
result of the S/G classification is available in the source lists via
the 2DPHOT flag. Flag values are: 1 (high-confidence star candidates),
2 (objects with FWHM smaller than stars in the stellar locus, e.g.,
some cosmic-rays and/or other unreliable sources), 4 (stars according
to S/G separation), and 0 otherwise (galaxies); flag values are
summed, so 2DPHOT = 5 signifies a high-confidence star candidate that
is also above the S/G separation line. Table
\ref{Tab:singlebandcolumns} lists all columns present in these source
lists.

\subsubsection{Multi-band catalogue}
\label{Sec:MultiBandCatalogue}

The multi-band catalogue delivered as part of KiDS-ESO-DR2 is intended as a "general purpose" catalogue and relies on
the double-image mode of {\sc SExtractor} and also
incorporates information obtained using the {\sc KiDS-CAT} software described
above. {\sc SExtractor} is run four times for each tile, using the $u$,
$g$, $r$ and $i$ KiDS-ESO-DR1/2 coadds as measurement images, to extract source
fluxes in each of the filters. The $r$-band coadd is used as detection
image in all runs, since it provides the highest
image-quality in almost all cases. For the same reason, several shape measurements are based only on the $r$-band data. 
In the future, it is foreseen that the catalogue S/G separation will make use of colour information and/or PSF modeling. Currently the S/G separation information included in the catalogue is the same as in the $r$-band source list.
The detection set-up is identical to that employed
for the source detection for the single-band source list
(Table \ref{Tab:SExtractorSetup}). Masking information is provided for all filters in the same fashion as in the single-band source lists. Compared to the single-band source lists the number of measured
parameters is reduced. Table \ref{Tab:MultiBandColumns} lists
the columns present in these source lists.

To account for seeing differences between filters aperture corrected fluxes are provided in the catalogue. The aperture corrections were calculated for each filter by comparing the aperture fluxes with the flux in a 30 pixel aperture, the aperture used for photometric calibration, and the aperture-corrected fluxes are included in the catalogue as separate columns. Source magnitudes and fluxes are not corrected for Galactic foreground or intergalactic extinction.

In order to prevent sources in tile overlaps to appear as multiple entries in the catalogue, the survey tiles have been cropped and connect seamlessly to one another. This results in slightly shallower data along the edges of the tiles (Fig. \ref{Fig:DithersFootprint}), similar to the areas partially covered by CCD gaps. Overall, all included areas are covered by at least three exposures. In future releases this will be improved by combining information from multiple tiles in overlap regions.

   \begin{figure*}[ht]
   \centering
   \includegraphics[width=\textwidth]{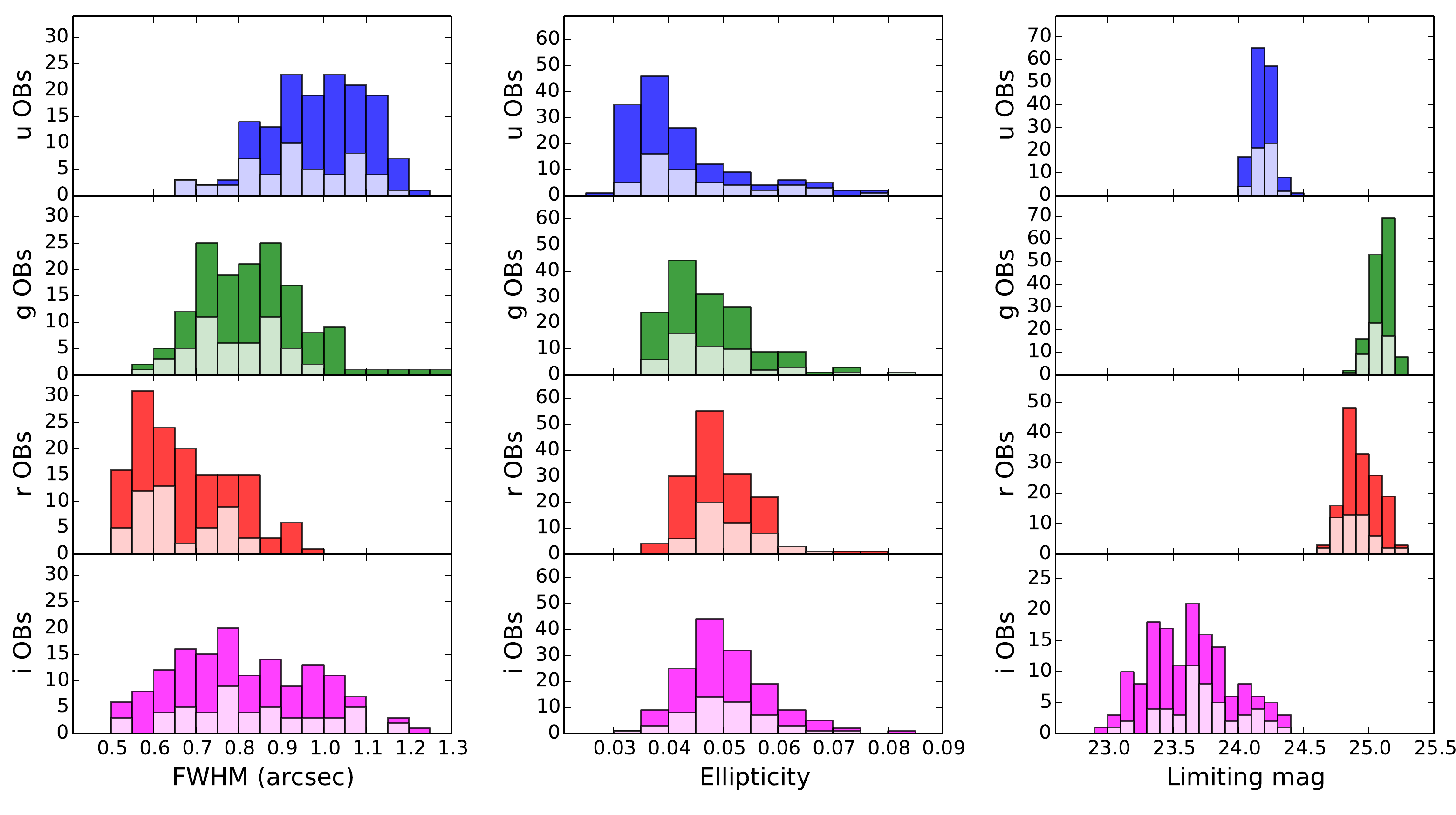}
   \caption{Data quality for KiDS-ESO-DR1 and KiDS-ESO-DR2. {\it Left:} average PSF size (FWHM)
distributions; {\it center:} average PSF ellipticity distributions; {\it right:} limiting magnitude distributions (5$\sigma$ AB in 2\arcsec\ aperture). The distributions are per filter: from top to bottom $u$, $g$, $r$, and $i$, respectively. The lighter portions of the histograms correspond to the 50 tiles in KiDS-ESO-DR1 and the darker portions to the 98 tiles in KiDS-ESO-DR2.}
              \label{Fig:DataQuality}
    \end{figure*}

\section{Data quality}
\label{Sec:DataQuality}

\subsection{Intrinsic data quality}
\label{Sec:IntrinsicDataQuality}

In Fig. \ref{Fig:DataQuality} the obtained seeing (FWHM), PSF
ellipticity, and limiting magnitude distributions per filter are
shown, to illustrate the raw data quality.  The PSF ellipticity is
defined here as $1-B/A$, where $A$ and $B$ are the semi-major and
semi-minor axis, respectively. Limiting magnitudes are 5$\sigma$ AB in
a 2\arcsec\ aperture and determined by a fit to the median SNR,
estimated by 1/MAGERR\_AUTO, as function of magnitude.  In case of the
filters observed in dark time ($u$, $g$, $r$) the FWHM distributions
reflect the different observing constraints, with $r$-band taking the
best conditions. Since $i$-band is the only filter in which the data
are obtained in bright time, it is observed under a large range of
seeing conditions. Average PSF ellipticities are always small: $<0.1$
(the average is over the absolute value of ellipticity, regardless of
the direction of ellipticity). The wide range of limiting magnitudes
in $i$-band is caused by the large range in moon phase and thus sky
brightness.

\subsubsection{Point-spread-function}
\label{Sec:PointSpreadFunction}

\begin{figure*}
   \centering
   \includegraphics[width=\textwidth]{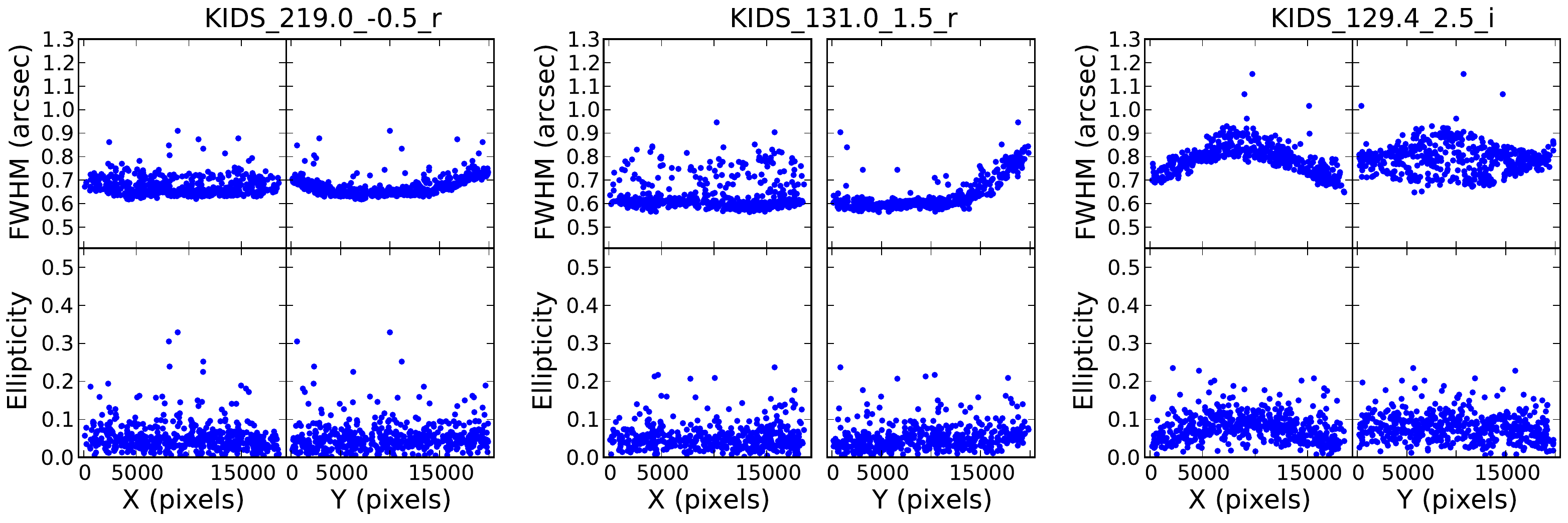}
   \caption{Examples of significant PSF size and ellipticity variation across three KiDS image stacks: KIDS\_219.0\_-0.5\_r (left), KIDS\_131.0\_1.5\_r (center), and KIDS\_129.4\_2.5\_i (right). In each subplot the upper panels show the FWHM in arcseconds and the lower panels the ellipticity, both plotted versus pixel coordinates X (left panels) and Y (right panels). Points correspond to the 500 brightest, unsaturated and unflagged point sources in each field.}
              \label{Fig:PSFvariations}
    \end{figure*}

Designed with optimal image quality over the full square degree field-of-view in mind, the VST/OmegaCAM system is capable of delivering images with an extremely uniform PSF. 
A typical example of the PSF ellipticity and size pattern for an observation taken with a nominal system set-up is shown in the left panel of Fig. \ref{Fig:PSFvariations}. 
Of course, as must be expected from a newly commissioned instrument, the set-up of the optical system is not always perfect and different imperfections can lead to a variety of deviations from a stable and round PSF. Most commonly encountered patterns are related to imperfect focus (increased PSF size either on the outside or on the inside of the FOV) and mis-alignment of the secondary mirror (increased PSF size and ellipticity along one edge of the FOV).

\begin{figure*}
   \centering
   \includegraphics[width=\textwidth]{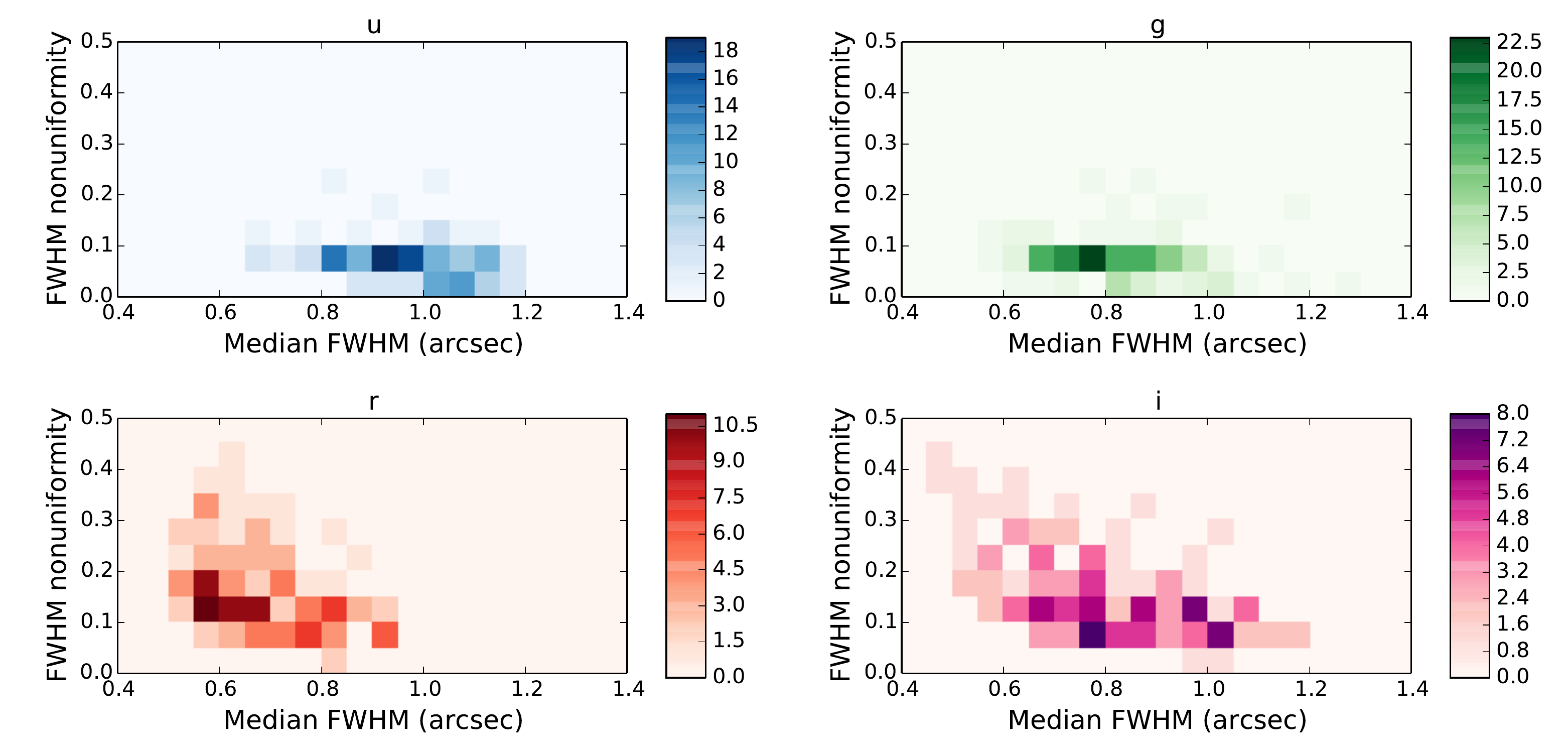}
   \caption{PSF size nonuniformity vs. median PSF size for KiDS-ESO-DR1/2 tiles, per filter ({\it top left:} u; {\it top right:} g; {\it bottom left:} r; {\it bottom right:} i). The nonuniformity is calculated as the PSF size (FWHM) difference between the 4 CCDs with the biggest PSF and the 4 CCDs with the smallest PSF within a coadd, divided by the average PSF size. The colour scale indicates the number of tiles in each bin.}
              \label{Fig:PSFsizes}
    \end{figure*}

To quantify the stability of the PSF we calculate a PSF size and ellipticity for every observation based on the PSF sizes in 32 regions in the coadded image that correspond roughly to the 32 CCDs. Systematic variations in ellipticity are usually less clear than in PSF size (see Fig. \ref{Fig:PSFvariations}), which is why the latter is used for monitoring PSF stability. The average PSF size in the 4 regions with the smallest PSF is subtracted from the average PSF size in the 4 regions with the largest PSF, and the result divided by the average PSF size of the whole image. The distribution per filter of this PSF size nonuniformity is shown versus the median FWHM in Fig. \ref{Fig:PSFsizes}. As the median PSF size (i.e. the seeing) increases, the nonuniformity drops, as any differences due to optical imperfections are smoothed out. However, even during very good seeing conditions the PSF size variation rarely exceeds 25\% and in most cases is around 10\%. 

\begin{figure*}
   \centering
   \includegraphics[width=\textwidth]{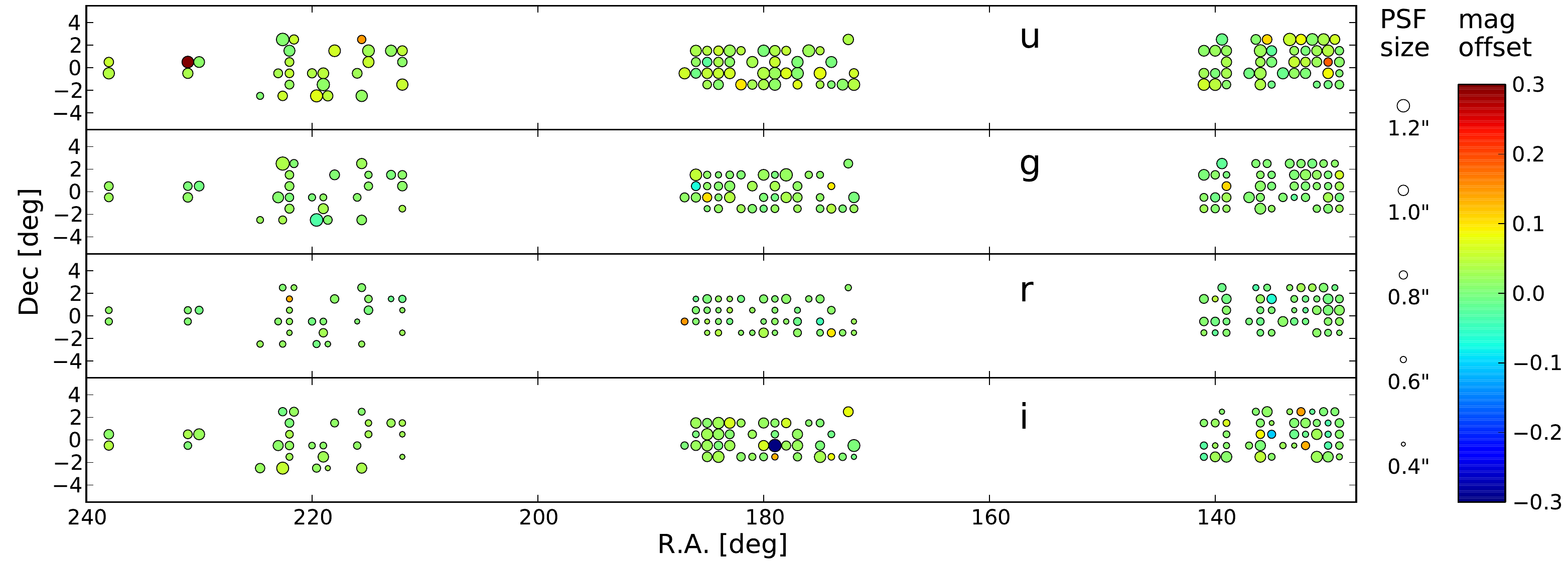}
   \caption{PSF sizes and photometric calibration for all tiles in KiDS-North as function of right ascension and declination. The average PSF size in each coadded image is indicated by the size of the circle, and the magnitude offset with respect to SDSS by its color. From top to bottom the panels correspond to the $u$, $g$, $r$, and $i$ filters.}
              \label{Fig:PhotRADEC}
    \end{figure*}

The median PSF sizes for all tiles in the KiDS-North field are indicated in Fig.~\ref{Fig:PhotRADEC}. Both in KiDS-North and KiDS-South there are no systematic gradients in PSF size over the survey area.

\subsection{Photometric quality}
\label{Sec:PhotometricQuality}

The matter of photometric quality can be divided in two parts, namely the uniformity of the photometry within each tile, and the quality of the photometric calibration per tile/filter.
Since the distribution of tiles included in KiDS-ESO-DR1/2 is not contiguous, with many
isolated tiles, a complete photometric homogenization of the
entire data set is impossible, and the photometric calibration is currently done per tile and per filter. The quality of the photometry will improve greatly in future releases when, using significant contiguous areas, a global calibration for the entire survey will be performed. 

\subsubsection{Comparison to SDSS}

Both the internal photometric homogeneity within a coadd and the quality of the absolute photometric
scale is assessed by comparing the KiDS photometry to SDSS DR8 \citep{aihara11}, which is photometrically stable to $\sim$1\% \citep{padmanabhan+08}. For this purpose, the aperture-corrected magnitudes in the multi-band catalogue were compared to PSF magnitudes of stars in SDSS DR8. Only unmasked stars with photometric uncertainties in both KiDS and SDSS smaller than 0.02 mag in $g$, $r$, and $i$ or 0.03 mag in $u$ were used. This comparison is only possible for all tiles in the KiDS-North field, but since KiDS-South was calibrated in the same way as KiDS-North, we expect the conclusions to hold for all data.

\begin{figure}
   \centering
   \includegraphics[width=\columnwidth]{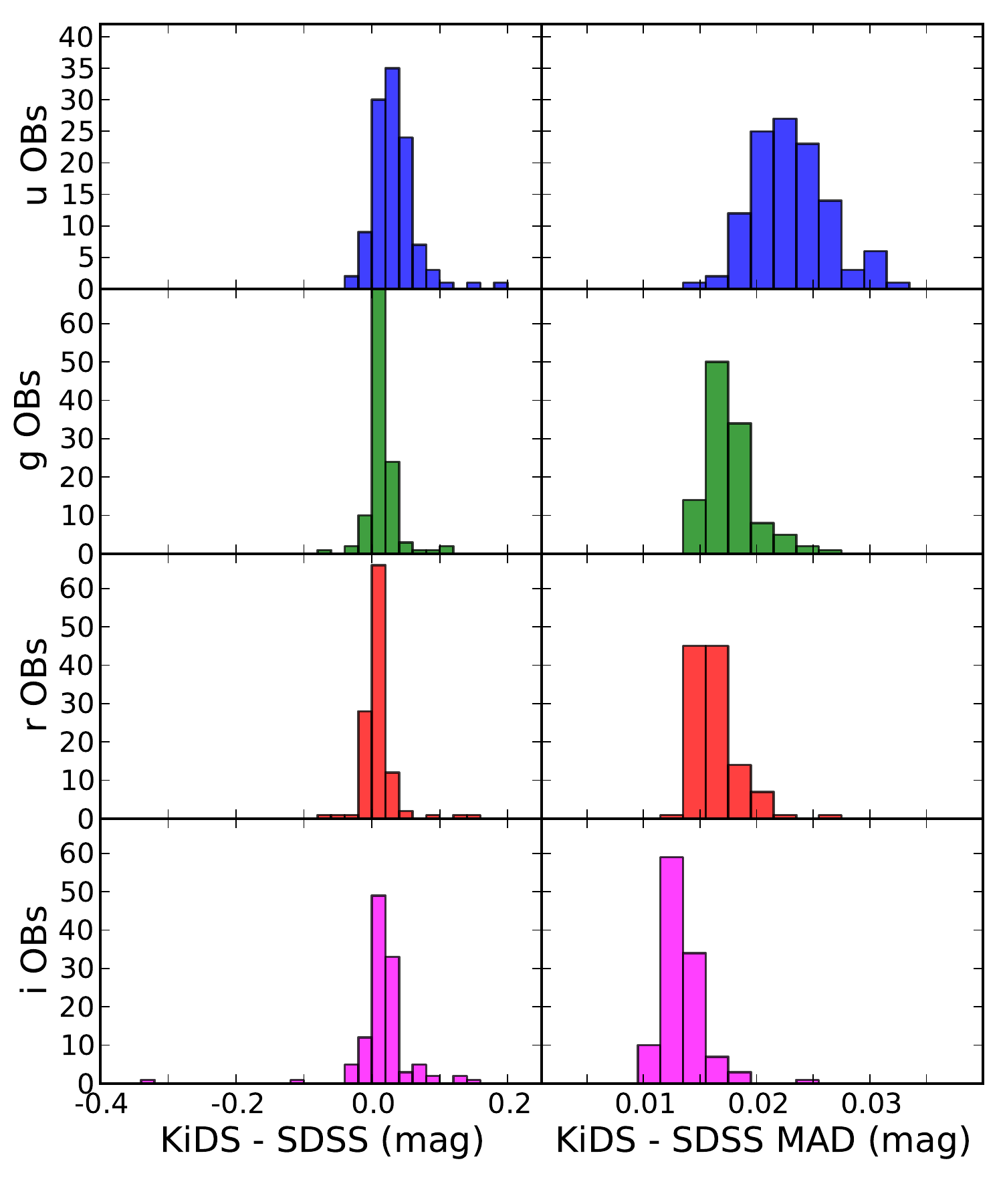}
   \caption{Comparison of KiDS aperture-corrected photometry to SDSS DR8 PSF photometry for stars in KiDS-North. Distributions of the median offsets are shown in the left column and the distributions of Median Absolute Deviations from these offsets in the right column. In both columns the subpanels correspond, from top to bottom, to $u$, $g$, $r$ and $i$.}
    \label{Fig:PhotometricQuality1}
    \end{figure}

The consistency of the photometric calibration is illustrated in Fig. \ref{Fig:PhotometricQuality1}, where histograms of the distributions of photometric offsets between KiDS and SDSS are shown for the overlapping tiles in KiDS-North.  A systematic offset of $\sim0.02$ mag is present in all filters, possibly due to the fact that nightly zero-points are determined using a fixed aperture on stars in the SA field without aperture correction. The scatter and occasional outliers are due to non-photometric conditions (during either KiDS or SA field observations) and, particularly in case of the $u$-band, use of default zero-points. In Fig.~\ref{Fig:PhotRADEC} the photometric offsets are plotted as function of RA and DEC, demonstrating that there are no large-scale gradients present.
All photometric offsets determined from this comparison with SDSS are available in the Source catalogue table on the KiDS website\footnote{\url{http://kids.strw.leidenuniv.nl/DR2}}.

\begin{figure}
   \centering
   \includegraphics[width=\columnwidth]{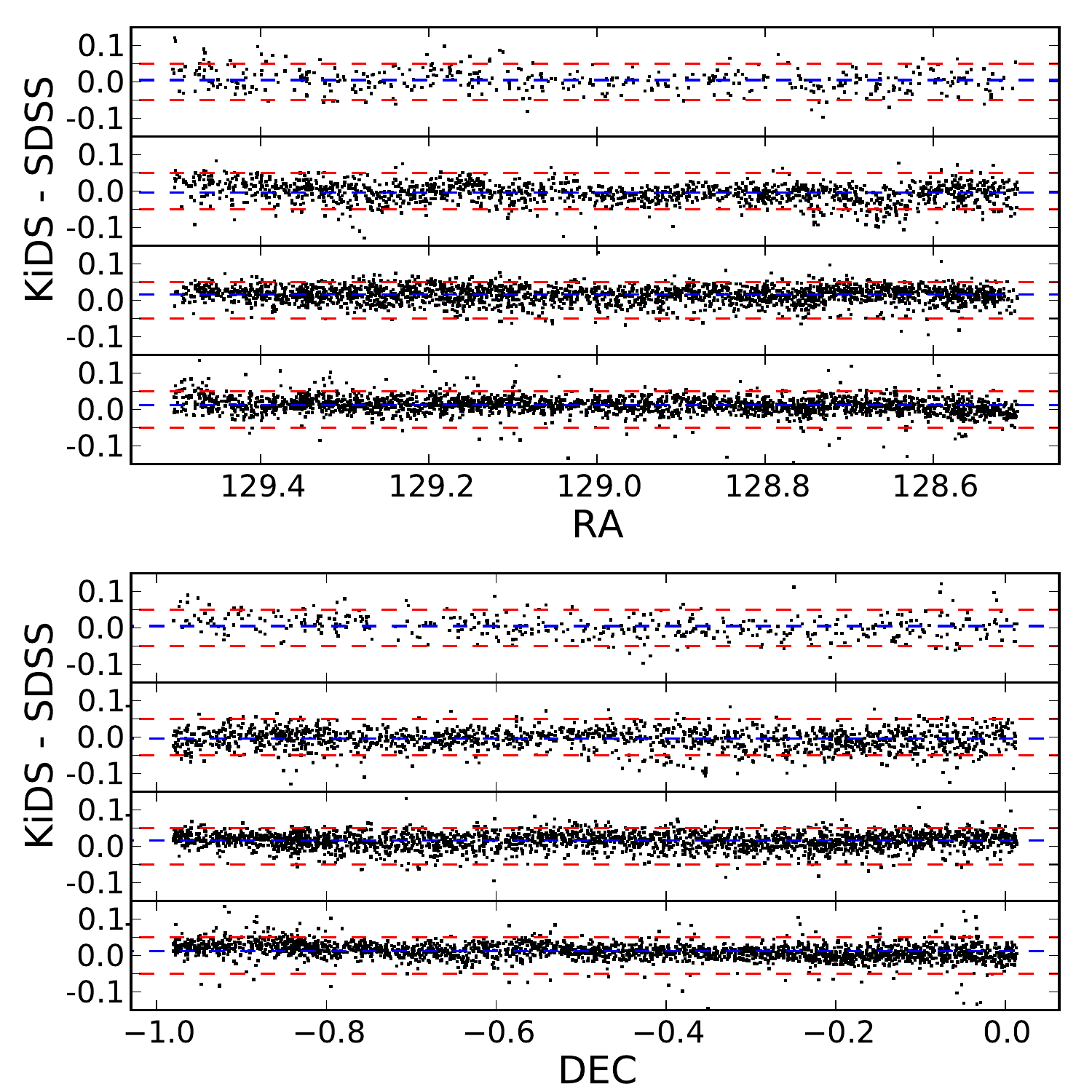}
   \caption{Photometric comparison between aperture-corrected KiDS photometry and SDSS DR8 PSF photometry for (unmasked) stars in tile KIDS\_129.0\_-0.5. {\it Top:} offsets versus right ascension. {\it Bottom: } offsets versus declination. Subpanels correspond, from top to bottom, to $u$, $g$, $r$ and $i$, respectively. Each dot corresponds to a star, with the average indicated by a blue dashed line and with red dotted lines indicating +0.05 and -0.05 magnitudes.}
    \label{Fig:SDSSComparison}
    \end{figure}

Figure \ref{Fig:SDSSComparison} shows the residuals between KiDS and SDSS DR8 magnitudes for one tile (KIDS\_129.0\_-0.5), which is a representative example. Generally speaking, the photometry in a filter within one tile is uniform to a few percent. The right column of the left panel in Fig. \ref{Fig:PhotometricQuality1} shows the distribution of the Median Absolute Deviation (MAD) of the stellar photometry between KiDS and SDSS for the tiles in KiDS-North, demonstrating the photometric stability within survey tiles. The relatively poor photometry in $u$-band is due to the lack of photometric homogenization within a tile.

   \begin{figure}
   \centering
   \includegraphics[width=\columnwidth]{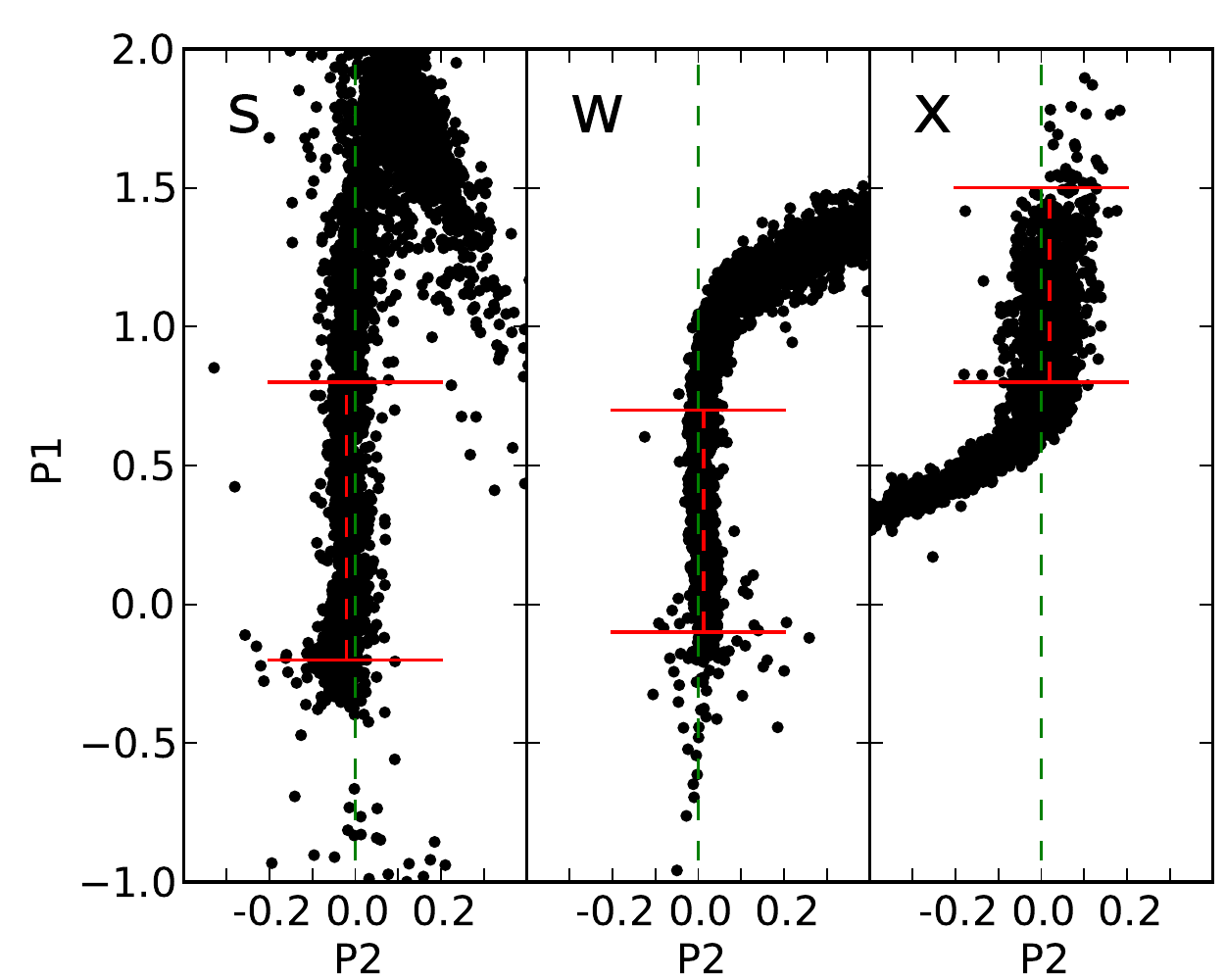}
   \caption{Stellar locus in ``principal colour'' space for tile KIDS\_129.0\_-0.5. For the definition of the three colour planes see \cite{ivezic04}. In SDSS filters the stellar locus should lie at $p2s = p2w = p2x = 0$, indicated by the green dashed line. The dashed red line shows the median stellar locus colour and the solid red lines the range in the $p1$ colours used for this analysis. Only unflagged stars with $r<21$ are plotted here.}
    \label{Fig:Principalcolours}
    \end{figure}

   \begin{figure}
   \centering
   \includegraphics[width=\columnwidth]{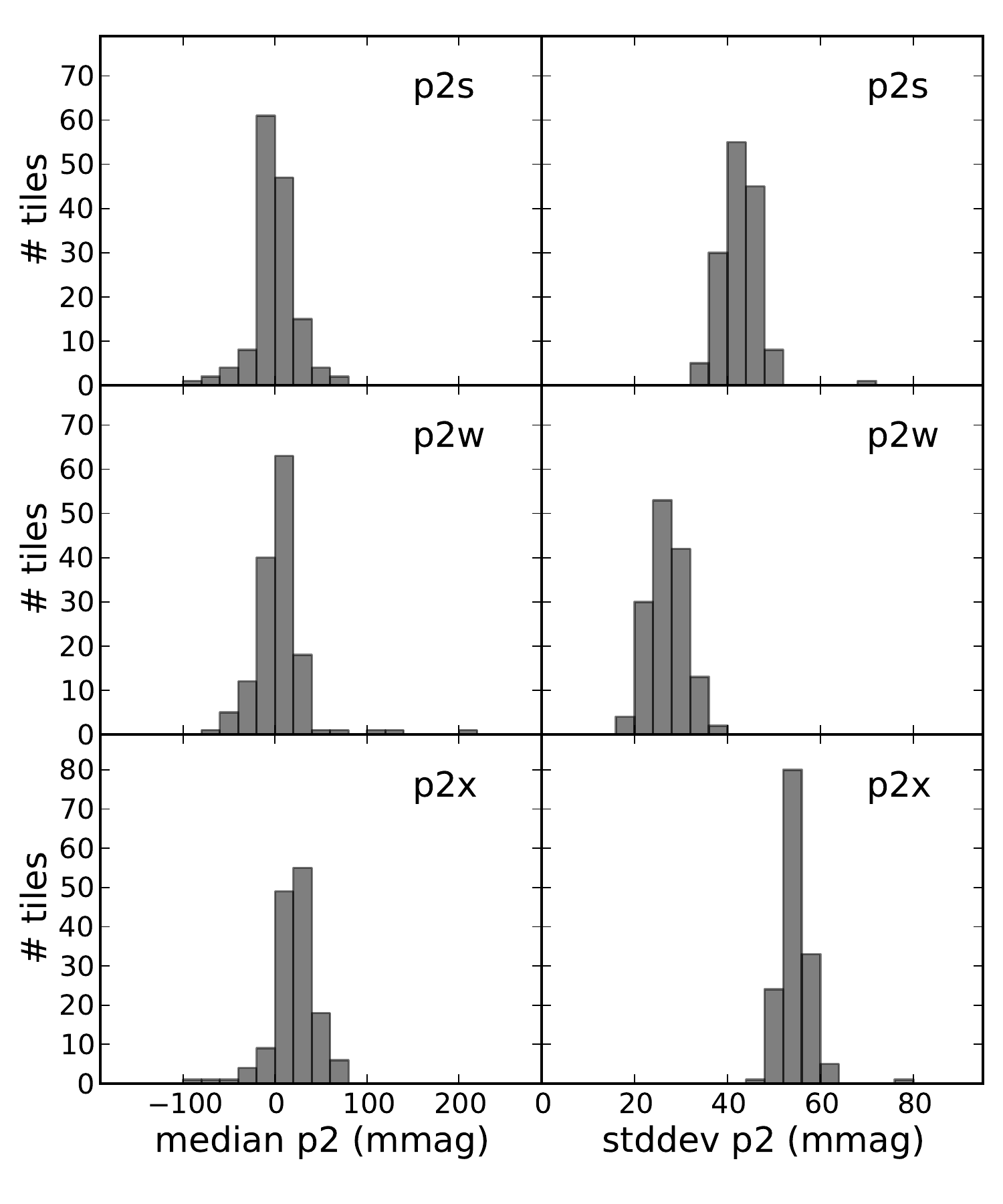}
   \caption{Stellar locus colour analysis using ``principal colours'' defined in \cite{ivezic04} and based on KiDS aperture-corrected stellar photometry for all tiles. Distributions of the offsets from the expected stellar locus colour are plotted in the left column. In the right column the distributions of the widths of the stellar locus (measured by standard deviation) are shown.}
    \label{Fig:PhotometricQuality2}
    \end{figure}

\subsubsection{Stellar locus}

A second quality assessment of the photometry is done by comparing stellar photometry to empirical stellar loci, in the form of the "principal colours" as defined by \cite{ivezic04} based on SDSS photometry. Only unmasked and unflagged stars with $r$<21 are used for this analysis. Although small colour terms exist between KiDS and SDSS the stellar loci based on SDSS are a powerful tool to verify the photometric stability over the currently released tiles. Once more complete sky coverage allows an overall photometric calibration of the KiDS data, accurate stellar loci in the KiDS filters will be derived.

Figure \ref{Fig:Principalcolours} shows the stellar locus for the tile KIDS\_129.0\_-0.5 in the three principal colour planes. These colour planes are different combinations of the $ugri$ filters and denoted by $s$, $w$ and $x$. The median $p2$ colours are calculated for each of $p2s$, $p2w$ and $p2x$ by choosing all stars within the indicated limits and after clipping all stars more than 200 mmag away from the initial median $p2$ colour. In Fig. \ref{Fig:PhotometricQuality2} the distributions of the median principal colours are shown, together with the typical width of the stellar locus, measured by the standard deviation. The narrow distributions indicate that the colour of the stellar locus is typically stable to within 20 mmag. Also the stellar locus width is always stable to within 10 mmag. The large width of the stellar locus in $p2x$ is caused by the fact that this part of the locus is made up of relatively faint stars with large photometric uncertainties.

In this analysis three tiles stand out that have a median $p2w$ colour of >100 mmag. Comparing these with the SDSS photometry as discussed above shows that one corresponds to the tile with the largest offset ($-$0.32 mag) in $i$ and two correspond to tiles with large offsets (>0.1 mag) in $r$.

\subsubsection{Tile overlaps}

\begin{figure}
   \centering
   \includegraphics[width=\columnwidth]{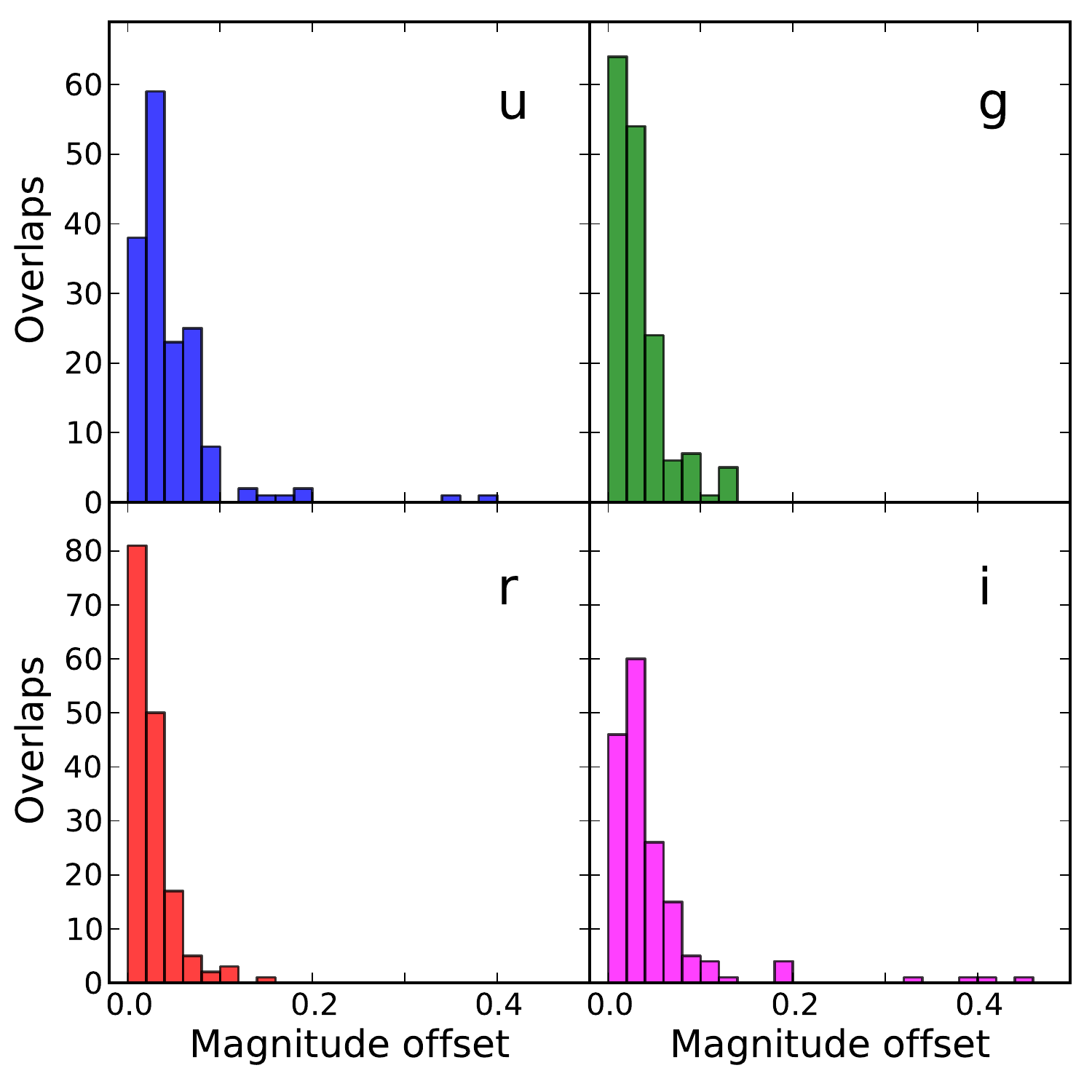}
   \caption{Tile-to-tile magnitude offsets in each filter measured from the overlap regions. In each available overlap region unflagged stars were selected brighter than 22 mag in the filter considered, and their MAG\_ISO magnitudes were compared, yielding a positive, mean offset.}
    \label{Fig:PhotometricQuality3}
    \end{figure}

Finally, we analyze the tile-to-tile photometric offsets directly from the KiDS data, using the overlapping areas between the tiles where available. This is done per filter, and only stars with a magnitude brighter than 22 in the respective filter are matched. In $u$ this results in a total number of sources per overlap region varying between 20 and 300, in $g$ and $r$ between 50 and 600 and in $i$ between 100 and 1,000. In some cases the PSF deteriorates at the edge of a tile, as described Sect.~\ref{Sec:PointSpreadFunction}, and the aperture-corrected magnitudes may be affected. For this reason MAG\_ISO magnitudes are used for this overlap analysis. The distributions of magnitude offsets in each filter, as determined from the tile overlaps, are shown in Fig.~\ref{Fig:PhotometricQuality3}. In the $gri$ filters the offsets are typicaly 0.02 mag, and for $u$ this increases to typically 0.04 mag. Taking into account that the photometry in the tile edges is of slightly poorer quality than in the center, due to the fact that these data are less deep, these values are in good agreement with the photometric comparison to SDSS.

\subsection{Astrometric quality}
\label{Sec:AstrometricQuality}

\begin{figure}
   \centering
   \includegraphics[width=\columnwidth]{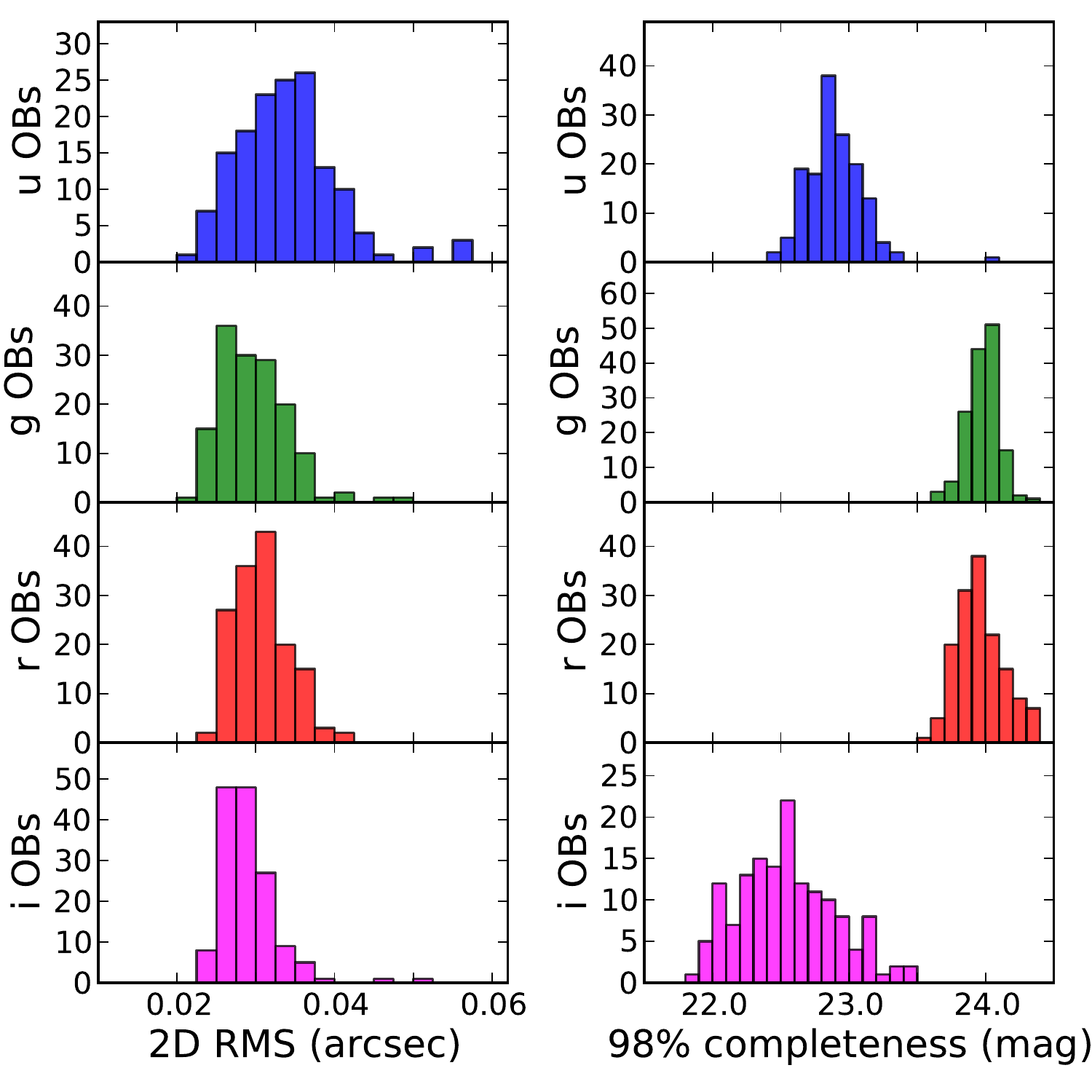}
   \caption{Overview of the astrometric quality and completeness of KiDS-ESO-DR1 and KiDS-ESO-DR2 (all data contained in the multi-band catalogue). {\it Left:} Histogram of the median relative astrometric offsets between the individual dithers and the corresponding coadd per survey tile. {\it Right:} 98\% completeness magnitude distributions for all tiles, based on the method of \cite{garilli99}. In both columns the panels correspond to $u$, $g$, $r$, and $i$ from top to bottom.}
    \label{Fig:AstrometryCompleteness}
    \end{figure}

The accuracy of the absolute astrometry ("KiDS vs 2MASS") is uniform over a coadd, with typical 2-dimensional (2-D) RMS of 0.31\arcsec\ in $g$, $r$, and $i$, and 0.25\arcsec\ in $u$, in line with expectations based on the fact that the majority of reference stars used is relatively faint. The lower RMS in $u$-band is most likely due to the fact that in this band on average brighter 2MASS sources are selected as reference sources. The accuracy of the relative astrometry ("KiDS vs KiDS"), measured by the 2-D positional residuals of sources between dithers, is also uniform across a single coadd. In Fig. \ref{Fig:AstrometryCompleteness} the accuracy of this relative astrometry of all coadds is shown. The typical 2-D RMS is $\sim0.03$\arcsec\ in all filters, but with a larger scatter in $u$ due to the smaller number of available reference sources.

\subsection{Completeness and contamination}
\label{Sec:CompletenessContamination}

Contamination of the multi-band catalogue by spurious sources was analyzed by means of a comparison of the overlap between KiDS and the CFHT Legacy Survey\footnote{\url{http://www.cfht.hawaii.edu/Science/CFHTLS/}}, the main deeper survey overlapping with the current data releases (CFHTLS-W2, using their final data release T0007\footnote{\url{http://terapix.iap.fr/cplt/T0007/doc/T0007-doc.html}}). For the analysis it is assumed that all KiDS sources not detected in CFHTLS-W2 are spurious. Since some fraction of real sources might be absent in the CFHTLS catalogues, the spurious fractions derived should be considered upper limits.

\begin{figure}
   \centering
   \includegraphics[width=\columnwidth]{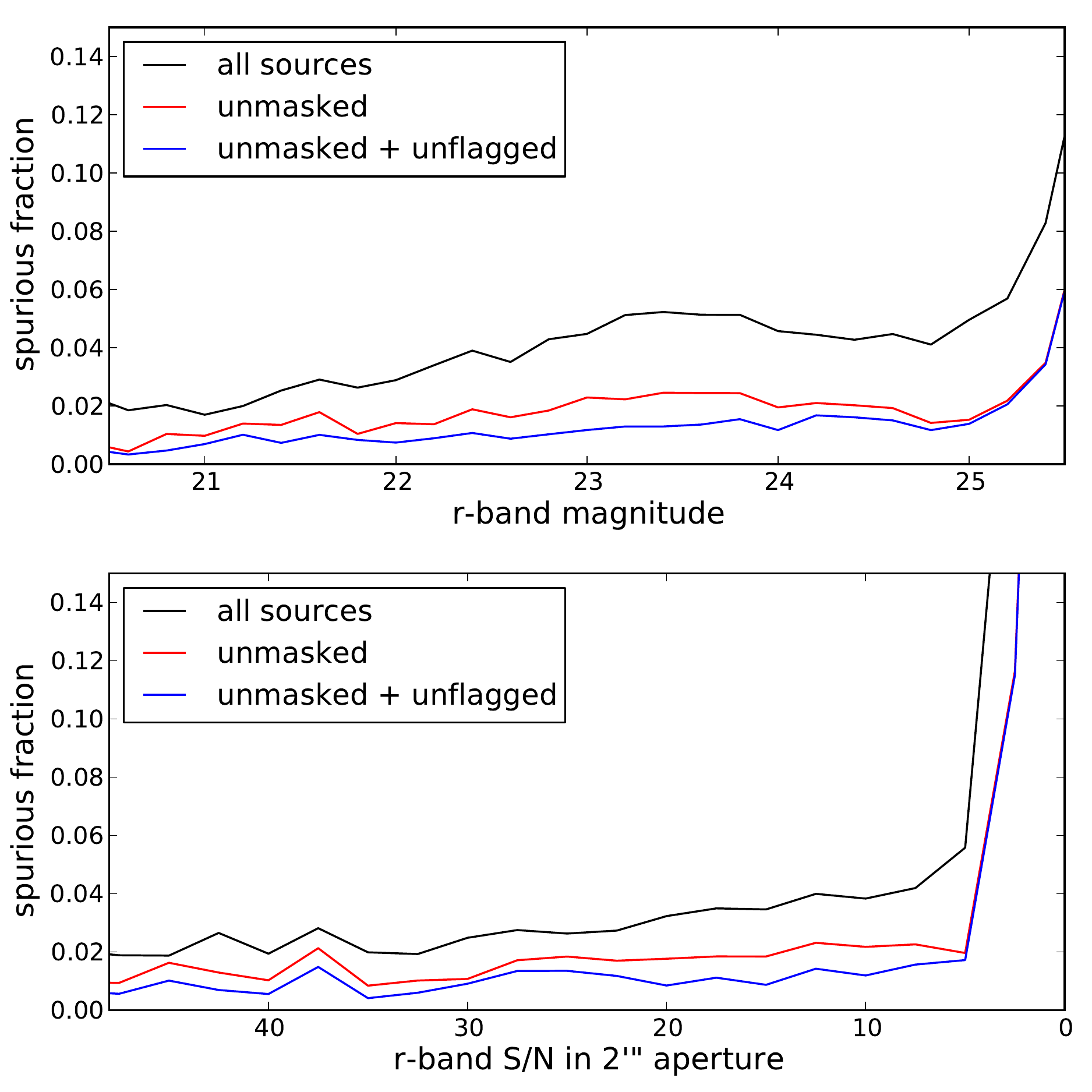}
   \caption{Spurious source contamination in the overlap between the KiDS multi-band catalogue and the CFHTLS-W2 field (this corresponds to the KiDS tiles KiDS\_135.0\_-1.5 and KiDS\_136.0\_-1.5). {\it Top:} spurious fraction vs. $r$-band magnitude (MAG\_AUTO). {\it Bottom:} spurious fraction vs. signal-to-noise in a 2\arcsec\ aperture. The black line corresponds to all sources, while the red line excludes sources in masked areas, and the blue line excludes sources in masked areas and sources with a non-zero {\sc SExtractor} flag.}
    \label{Fig:Contamination}
    \end{figure}

Figure \ref{Fig:Contamination} shows the spurious fractions derived from this comparison as function of magnitude ($r$-band MAG\_ISO) and signal-to-noise (in a 2\arcsec\ aperture). When all sources in the catalogue are considered the fraction of spurious sources is estimated to be <5\% down to a very low SNR of $\sim5$ within a 2\arcsec\ aperture. Filtering sources based on masking information reduces this fraction to $\sim2\%$, demonstrating that caution is required when using faint sources in masked regions. When sources are filtered both on masking information as well as {\sc SExtractor} detection flags, the spurious fractions drops even further to $\sim1\%$, yielding a very clean catalogue down to the detection limit.

An internal estimate of the completeness for the KiDS data is provided per tile, based on the method of \cite{garilli99}. It determines the magnitude at which objects start to be lost in the source list because they are below the brightness threshold in the detection cell. The implementation is similar to \cite{labarbera10}.  Estimates of the completeness obtained by comparison to deeper CFHTLS-W2 data are consistent with these internally derived values. The distributions of the 98\% completeness magnitudes for all tiles are shown in Fig. \ref{Fig:AstrometryCompleteness}. Comparison with Fig. \ref{Fig:DataQuality} shows that the 98\% completeness limits are typically $\sim1$ magnitude brighter than the limiting magnitude for $g$, $r$ and $i$ and $\sim1.3$ magnitudes brighter in $u$. For the completeness of the multi-band catalogue the values for the $r$-band of each tile apply. 

\subsection{Data foibles}
\label{Sec:DataFoibles}

   \begin{figure*}
   \centering
   \includegraphics[width=14cm]{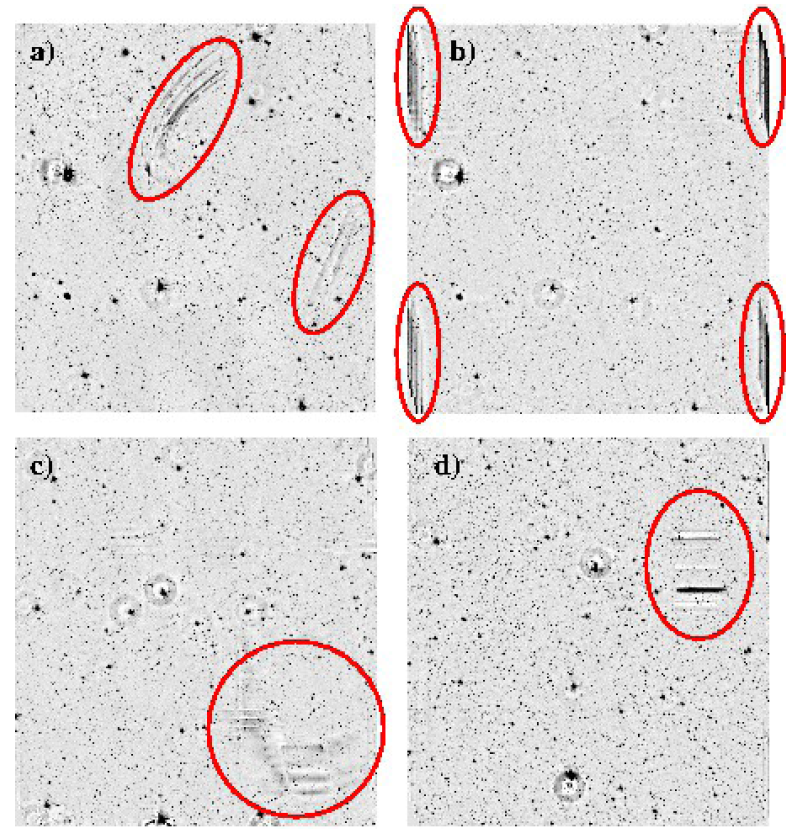}
   \caption{Examples of known issues in the KiDS-ESO-DR1/2 VST/OmegaCAM data, highlighted by the red ellipses. a)
Light patterns caused by reflections and scattered light of bright sources outside the FOV. b) Vignetting
and scattering by CCD masks at the corners of the CCD array. c) Extended background
structures caused by scattering of moonlight. d) Patterns caused by defective video board of CCD 82. Improved telescope baffles installed in early 2014 significantly reduce the occurrence of scattered light in future data releases.}
    \label{Fig:Foibles}
    \end{figure*}

\subsubsection{Scattered light and reflections}

Some of the main challenges in the analysis of early VST/OmegaCAM data are related to scattered light and reflections.
Due to the open structure of the telescope, light from sources outside the field-of-view
often affects the observations. This expresses itself in a number of ways:
\begin{itemize}
\item reflections: in some cases strong reflected light patterns are seen in the focal plane; these
are caused by light from bright point sources outside the field-of-view and can occur in
all filters. Some examples are shown in Fig. \ref{Fig:Foibles}a.
\item vignetting by CCD masks: vignetting and scattering by the masks present at the corners
of the focal plane array, and at the gaps between the rows of CCDs; this effect is particularly
strong in i-band due to the bright observing conditions. The effect near the CCD gaps is largely
corrected for, but in many cases the areas in the corners of the CCD array is strongly affected.
Examples are shown in Fig. \ref{Fig:Foibles}b.
\item extended background artifacts: related to the reflections mentioned before, this is mostly
seen in i-band and probably caused by moonlight. An example is shown in Fig. \ref{Fig:Foibles}c.
Most of these effects are not (yet) corrected for in the current data processing, but strongly affected
regions are included in the image masks and affected sources are flagged in the source lists and catalogue.
\end{itemize}
Improvements to the telescope baffles that were installed in early 2014 significantly improve scattered light suppression.

\subsubsection{Individual CCD issues}

There are two issues related to individual CCDs that noticeably affect this data delivery:
\begin{itemize}
\item CCD 82: this CCD suffered from random gain jumps and related artifacts until its video
board was replaced on June 2 2012. Artefacts as shown in Fig. \ref{Fig:Foibles}d are sometimes visible
in the image stacks due to this problem. Photometry in this CCD can be used due to the
cross-calibration with neighbouring CCDs in the dithered exposures, but part of the CCD
is lost. These features are included in the image masks and affected sources are flagged.
\item CCD 93: during a few nights in September 2011 (Early Science Time) one CCD was effectively
dead due to a video cable problem. One observation included in this data delivery
does not include this CCD: the $i$-band observation of KIDS\_341.2\_-32.1.
\end{itemize}

\section{First scientific applications}
\label{Sec:ScienceResults}

The Kilo-Degree Survey was designed for the central science case of
mapping the large-scale matter distribution in the Universe through
weak gravitational lensing and photometric redshifts. Several other
science cases were also identified from the early design stages
including studying the structure of galaxy halos, the evolution of
galaxies and galaxy clusters, the stellar halo of the Milky Way, and
searching for rare objects such as high-redshift QSOs and strong
gravitational lenses. While for the main science goal of KiDS the full
survey area is required, scientific analyses focusing on the other
science cases are currently already on-going. Below we demonstrate the
quality and promise of KiDS by means of these on-going research
efforts.

\subsection{Photometric redshifts and weak gravitational lensing}
\label{Sec:EarlyWL}

An analysis of the weak gravitational lensing masses of galaxies and
groups in the KiDS images is one of the main early scientific goals of
the survey \citep{viola15,vanuitert15,sifon15}. Such an analysis requires two types of
measurements from the KiDS data: galaxy shapes for shear measurements,
and galaxy colours for photometric redshifts. The shapes are measured
with a dedicated pipeline based on the CFHTLenS analysis
\citep{heymans+12,erben+13,miller+13} that combines information from
individual exposures and avoids regridding of the pixels to maintain
image fidelity as much as possible. The colours and photometric
redshifts are derived from the {\sc BPZ} code \citep{benitez00,coe+06}
applied to the output from a PSF-homogenized photometry pipeline --- again 
an evolution from the CFHTLenS analysis pipeline, see
\cite{hildebrandt+12} --- which runs on the calibrated stacked images
released in KiDS-ESO-DR1/2. Further details of these dedicated analyses are
presented in \cite{techpaper-lensing-1}.

\begin{figure*}
   \centering
   \includegraphics[width=\textwidth]{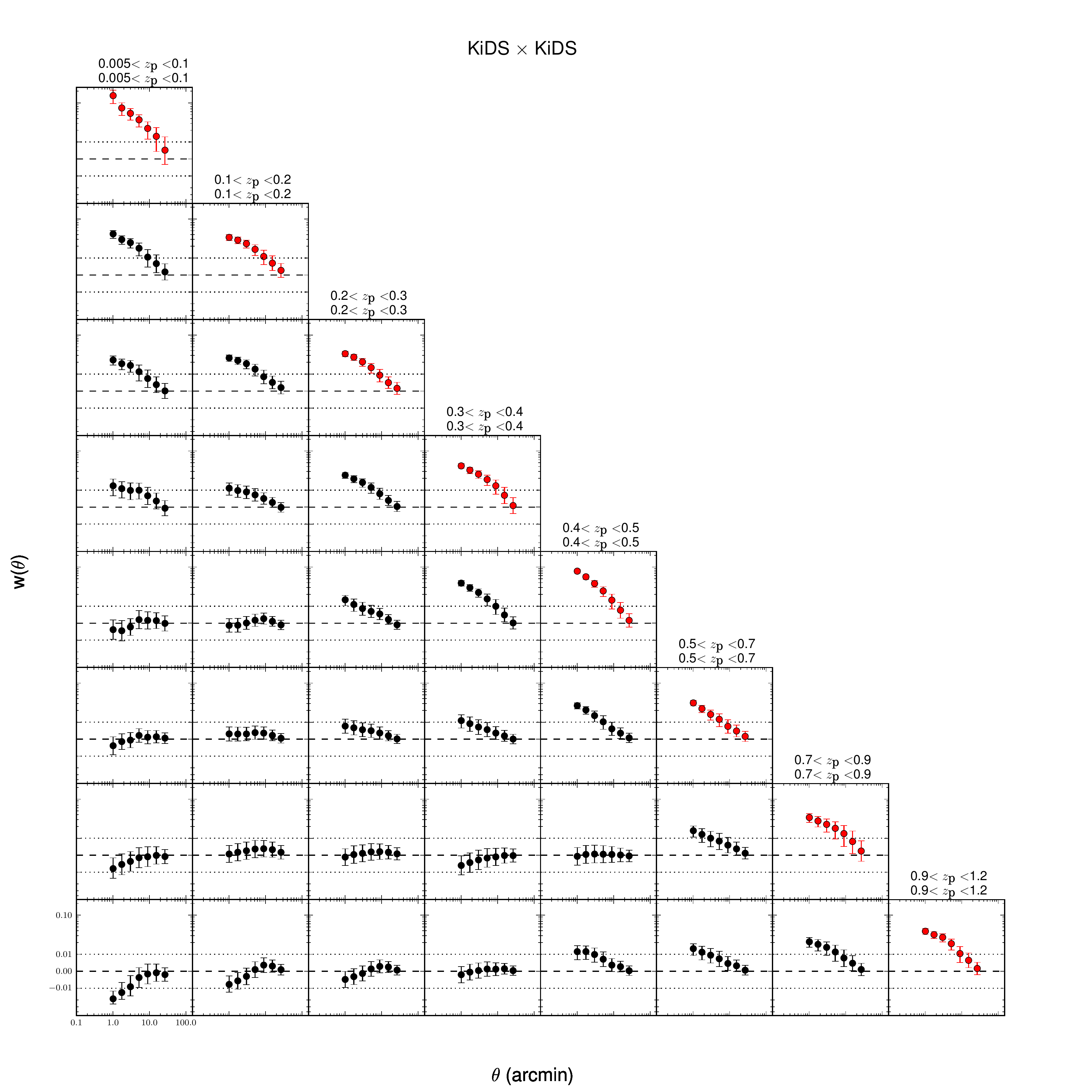}
   \caption{Angular (cross-)correlation of galaxy positions in different photometric redshift bins in the KiDS-ESO-DR1/2 fields, based on photometric redshifts derived with the BPZ code.  Note that the dotted lines at $\pm 0.01$ demarcate a band within which the y-axis is linear.  Only unmasked galaxies with extinction-corrected magnitudes brighter than $r=24$ and IMAFLAGS\_ISO=1 are included.  The errors are obtained from jackknife resampling, with each pointing being a jackknife sub-sample.  The strongest signals are seen along the diagonal panels and drop off in amplitude for more widely separated photometric redshift bins, which gives confidence in the quality of the photometric redshifts as well as the underlying photometry.}
    \label{Fig:wkidskids}
    \end{figure*}

As an illustration of the quality of the photometric redshifts, in
Fig.~\ref{Fig:wkidskids} we show the angular cross-correlations
$w(\theta)$ of the positions of galaxies in different photometric
redshift bins, on scales between 1 and 30 arcminutes
\citep{erben+09,benjamin+10}. We use {\sc Athena}
\citep{athena} to calculate w($\theta$) using the
\citet{landy+93} estimator.  Errors are obtained from
jackknife resampling, with each pointing being a jackknife
sub-sample. The figure shows a clear clustering signal of galaxies
within the same redshift bin (panels on the diagonal). Most
off-diagonal panels show a smaller cross-correlation amplitude than
the corresponding auto-correlations, and mostly this signal is seen
only between neighbouring redshift bins, as would be caused by scatter
of the photometric redshift estimates into neighbouring bins. The fact
that no strong signal is seen further away from the diagonal shows
that the level of catastrophic failures in the redshifts is low, and
provides confidence in the photometric redshifts as well as the
underlying photometry reported here. 

This photometric redshift-only cross-correlation check is
complementary to the spectroscopic redshift-photometric redshift
cross-correlations presented in \cite{techpaper-lensing-1}.  The latter
analysis has the advantage of utilising spectroscopic redshifts, which
provide a better representation of the 'absolute truth'; however, the
spectroscopic redshifts only reach up to $z \sim 0.5$, so there is
additional information provided by the photometric redshift-only
cross-correlations out to $z \sim 1$.  We refer the interested reader
to \cite{techpaper-lensing-1} for details of the cross-correlation
analysis between the photometric redshifts and the available
spectroscopic redshifts.

\subsection{Photometric redshifts from machine learning}
\label{Sec:PhotzML}

Apart from the photometric redshifts described in Sect. \ref{Sec:EarlyWL}, 
photometric redshifts are also derived from KiDS-ESO-DR1/2 $ugri$ photometry using
the supervised machine learning model MLPQNA: a Multi-Layer Perceptron
feed-forward neural network providing a general framework for
representing nonlinear functional mappings between input and output
variables. QNA stands for Quasi Newton Algorithm, a variable metric
method used to solve optimization problems \citep{davidon1991} that,
when implemented as the learning rule of a MLP, can be used to find
the stationary (i.e. the zero gradient) point of the learning
function. The QNA implemented here is the L-BFGS algorithm from
\cite{shanno1970}.  Supervised methods use an extensive set (the
knowledge base or KB) of objects for which the output (in this case
the redshift) is known a-priori to learn the mapping function that
transforms the input data (in this case the photometric quantities)
into the desired output.  Usually the KB is split into three different
subsets: a \textit{training set} for training the method, a
\textit{validation set} for validating the training in particular
against overfitting, and a \textit{test set} for evaluating the
overall performance of the model \citep{cavuoti2012, brescia2013}.  In
the method used here the validation is embedded into the training
phase, by means of the standard leave-one-out k-fold cross validation
mechanism \citep{geisser1975}. Performances are always derived
blindly, i.e. using a test set formed by objects which have never been
fed to the network during either training or validation.  The MLPQNA
method has been successfully used in many experiments on
different data sets, often composed through accurate
cross-matching among public surveys (SDSS for galaxies:
\citealt{brescia2014}; UKIDSS, SDSS, GALEX and WISE for quasars:
\citealt{brescia2013}; CLASH-VLT data for galaxies:
\citealt{biviano2013}).

The KB of spectroscopic redshifts was obtained by merging
the spectroscopic datasets from GAMA data release 2 \citep{gamadr2} and
SDSS-III data release 9 \citep{sdssdr9}, while for the KiDS photometry
two different aperture magnitudes were adopted. The final KB includes
the optical magnitudes ($ugri$) within 4\arcsec and 6\arcsec diameters and
SDSS and GAMA heliocentric spectroscopic redshifts. GAMA redshifts
come with the normalized quality flag NQ.

\begin{figure}
\centering
\includegraphics[width=\columnwidth]{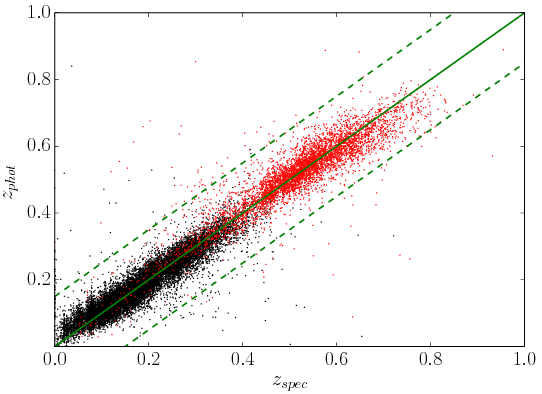}
\caption{The $z_{\rm{phot}}$ vs $z_{\rm{spec}}$ scatter plot of the blind test set KiDS-based optical photometric redshifts against the GAMA (black dots) and SDSS (red dots) spectroscopic redshifts, where the photometric redshifts have been obtained with the MLPQNA model. The dotted lines are the borders delimiting the catastrophic outlier sub-regions (i.e. including objects with $|\Delta z\_norm| > 0.15$).}
\label{Fig:photzML}
\end{figure}

A training set and a test set were created by splitting the KB into two
parts of 60\% and 40\%, respectively.  With these data sets two
experiments were performed, one using only the GAMA high-quality (HQ,
$NQ>2$) spectroscopic redshifts and one using a mix of GAMA and SDSS
spectroscopic information. These experiments are illustrated in
Fig.~\ref{Fig:photzML}.  To quantify the quality of the results we use the standard normalized photometric redshift error $\Delta z_{\rm{norm}}$ defined as
$|z_{\rm{spec}}-z_{\rm{phot}}|/(z_{\rm{spec}}+1)$. 
We find a $1\sigma$ scatter in $\Delta z_{\rm{norm}}$ of 0.027 and 
0.031 and a fraction of catastrophic outliers ($|\Delta
z_{\rm{norm}}|>0.15$) of 0.25\% and 0.39\% for the high-quality GAMA and
the GAMA+SDSS spectroscopic data sets, respectively.  The mixture of
GAMA HQ + SDSS spectroscopic data slightly extends the KB to higher
redshifts, but since there is still very limited information at
$z_{\rm{spec}} > 0.45$, this does not significantly increase the
performance at higher redshift. Finally, the presence of objects at
the minimum $z_{\rm{spec}} = 0$ indicates a residual presence of stars
within the sample. 
Further information about the experiments, results and the produced
catalogue of photometric redshifts is reported in
\cite{cavuoti+15}. A detailed comparison with redshifts from
SED-fitting will be presented in a forthcoming paper (Cavuoti et
al. in preparation).

\subsection{Galaxy structural parameters and scaling relations}
\label{Sec:GalaxyParameters}

Galaxies are the building blocks of the Universe and within the
framework of hierarchical structure formation they form bottom-up,
with smaller objects forming first, then merging into massive
structures (e.g.  \citealt{deLucia+06};
\citealt{Trujillo+06}). However, high-mass galaxies seem to have
formed most of their stars in earlier epochs and over a shorter time
interval than the lower mass ones (``downsizing'' scenario,
\citealt{Fontanot+09}). Characterising the properties of the luminous
matter and the way this has been assembled into dark-matter haloes
is crucial to reconcile theory and observations. Making use of the high spatial resolution, depth and area coverage of KiDS we aim to measure, across different redshift slices:
\begin{itemize}
\item total luminosity and the stellar mass through SED fitting (e.g. Le PHARE, \citealt{LePhare});
\item stellar population properties (\citealt{bruzual03}); 
\item structural parameters and colour gradients.
\end{itemize}
This will be one of the first samples of this depth over such a large area and wavelength 
coverage, allowing us to extend previous analyses (e.g. \citealt{SPIDER-I}) to higher redshifts.

\begin{figure}
\centering
\includegraphics[width=\columnwidth]{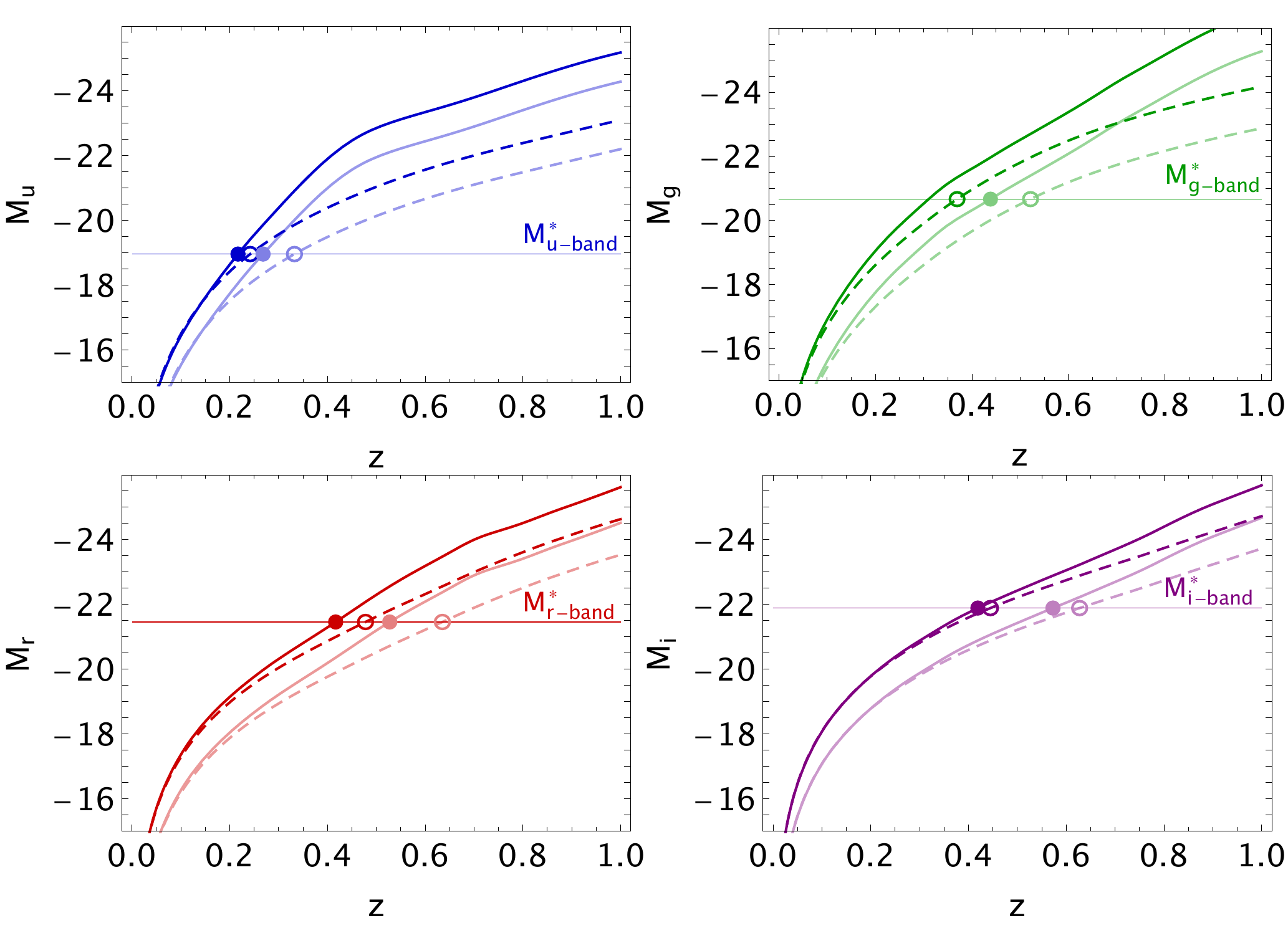}
\caption{90\% completeness of the photo-z (lighter colours) and high-SNR (darker colours) galaxy samples in absolute magnitude vs redshift.
Solid and dashed lines are for K-corrections from Elliptical and Scd models. 
Horizontal lines indicate $M^{\star}$ from the luminosity function analysis 
in \citet{Blanton+05_LF}.}
\label{fig:compl_vs_z}
\end{figure}

From the KiDS-ESO-DR1/2 catalogues we select galaxies based on the S/G separation discussed in Sect. \ref{Sec:SourceExtraction} with additional selection criteria on size (to further reduce contamination by stars), and remove any flagged or masked objects. For KiDS-ESO-DR1/2 this results in a sample of $\sim6.5$ million galaxies. For $\sim1$ million of these we have measured photometric redshifts as discussed in Sect. \ref{Sec:PhotzML}.
For the structural parameter derivation we have considered only galaxies with high $r$-band SNR (SNR$_r>50$) to reliably perform the surface brightness analysis (\citealt{labarbera08,SPIDER-I}). This sub-sample consists of around 350,000 galaxies.
The completeness of the whole sample has been discussed in Sect. \ref{Sec:CompletenessContamination}. For the sample with photometric redshifts the 98\% completeness magnitudes (all magnitudes used here are MAG\_AUTO) are $u$=22.3, $g$=22.1, $r$=20.4 and $i$=19.7, which correspond to 90\% completeness magnitudes of $u$=23.1, $g$=23.0, $r$=22.1 and $i$=21.2.
The stringent cuts on SNR for the sample with structural parameters results in 98\% completeness magnitudes of $u$=21.4, $g$=21.1, $r$=20.1, $i$=19.3 and 90\% completeness magnitudes of $u$=22.2, $g$=21.7, $r$=21.0, $i$=20.2.
These completeness magnitudes can be converted to absolute magnitudes as $M = m - DM(z_{\rm{phot}}) - K_{\rm{corr}}$, where $DM(z_{\rm{phot}})$ is the distance modulus based on the photometric redshift and $K_{\rm{corr}}$ is the K-correction, which is derived for two empirical models (elliptical and Scd galaxy). For these calculations a standard cosmology with $\Omega_{\rm{m}}=0.3$, $\Omega_{\Lambda} = 0.7$ and $H_{\rm 0} = 75$ km$\,$s$^{-1}$ Mpc$^{-1}$ is used, and the galactic foreground extinction is corrected based on the dust maps from \cite{SFD}. Figure \ref{fig:compl_vs_z} shows how the 90\% completeness limits vary with absolute magnitude as a function of redshift. Comparing this to the $M^{\star}$ from the luminosity function of low-$z$ SDSS galaxies (\citealt{Blanton+05_LF}) we find that $M^{\star}$ 
is reached at $z\sim$0.22, 0.31, 0.42, and 0.42 in our high-SNR galaxy sample and $z\sim$0.27, 0.44, 0.53, and 0.57 for the photo-z sample, in $u$, $g$, $r$ and $i$, respectively, if the Elliptical model is adopted. Slightly larger redshifts are found for late-type galaxies (see Fig.~\ref{fig:compl_vs_z}).

\begin{figure}
\centering
\includegraphics[width=\columnwidth]{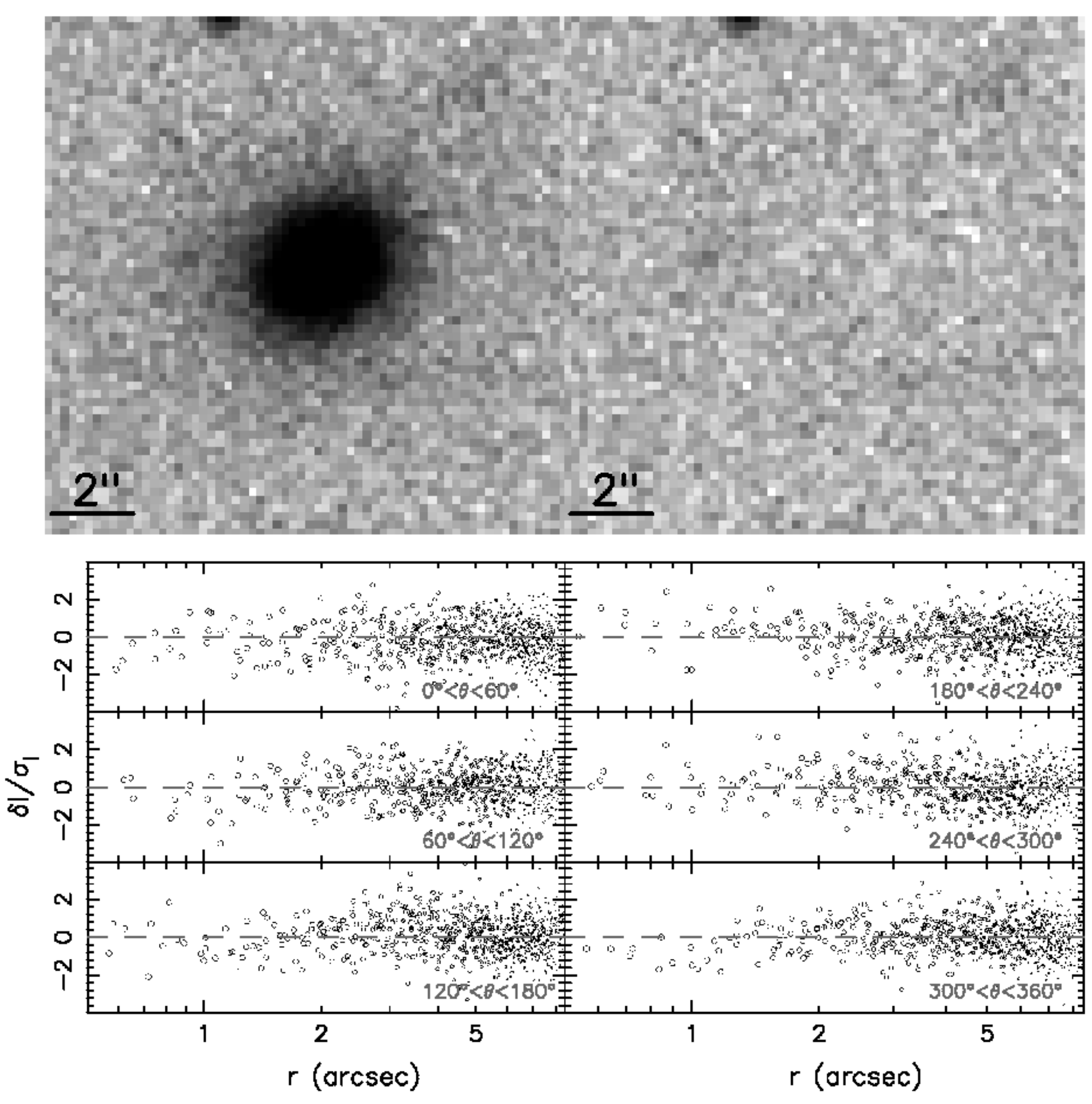}
\caption{Example of galaxy surface photometry fitting using 2DPHOT. {\it Top:} the thumbnail of an original KiDS $r$-band image with a galaxy in the center is shown on the left, and on the right the same thumbnail after subtracting the best-fit model. {\it Bottom:} normalized residuals after model subtraction, $\delta I/\sigma_I$, are plotted as function of the distance from the galaxy center. Each subpanel corresponds to a different bin of the polar angle, $\theta$, measured counterclock-wise from the positive x-axis. The residuals, normalized by the noise expected from the model in each pixel, $\sigma_I$, are remarkably consistent with zero for all positions in the image, implying a reduced $\chi^2 $ of about one.}
\label{fig:2Dphot}
\end{figure}

We make use of {\sc Galfit} (\citealt{galfit}) and {\sc 2DPHOT} (\citealt{labarbera08}) 
to fit PSF convolved S\'ersic profiles to the surface photometry and infer
the structural parameters (S\'ersic index, $n$, effective radius
$R_{\rm e}$, axial ratio, $q$, disky/boxy coefficient $a_4$, disk/bulge separation). In Fig. \ref{fig:2Dphot} we show an example of the results obtained with {\sc 2DPHOT}. 
The relationship between $R_{\rm e}$ or S\'ersic index and mass, luminosity and its evolution with
redshift has been demonstrated to be a fundamental probe of galaxy evolution and the role of mass accretion in galaxy merging (\citealt{Trujillo+07}; \citealt{Hilz+13}; \citealt{Tortora+14}). This characterization of 2D light profiles also allows one to determine
the colour gradients (\citealt{LaBarbera_deCarvalho09};
\citealt{Tortora+10CG}; \citealt{Tortora+11MtoLgrad};
\citealt{LaBarbera+11_CG}; \citealt{TNgenv12};
\citealt{LaBarbera+12_SPIDERVII_CG}), which can be compared to simulations (e.g.\citealt{Tortora+11CGsim};
\citealt{Tortora+13_CG_SIM}).

The resulting data set is introduced in \cite{Tortora+15} and applied to a first census of compact galaxies, a special class of objects, relic remnants of high-z red nuggets, which can provide significant constraints on the galaxy merging history.
The full analysis of the galaxy sample discussed above will be presented in Napolitano et al. (in prep).

\subsection{Galaxy cluster detection}
\label{Sec:ClusterDetection}

The evolution of the number density of massive galaxy clusters is an
important cosmological probe \citep[e.g.][]{allen+11}, particularly
at high redshifts ($z>0.5$). It is therefore important to increase the
number of known galaxy clusters.  The number of clusters as a function
of mass and redshift detected in a KiDS--like survey was evaluated
using the mock catalogues derived by \citet[][H12
hereafter]{2012MNRAS.421.2904H}, which are based on the semi--analytic
galaxy models built by \citet{2011MNRAS.413..101G} for the Millennium
Simulation \citep[MS; ][]{2005Natur.435..629S}. In particular, these
mock catalogues provide SDSS $ugri$ photometry for galaxies in 24 light
cones, 1.4 $\times$1.4 deg$^2$ each, as well as the $M_{200}$ mass of
their parent halos. Galaxy clusters were identified in the mock
catalogues following the recipe by \citet[][see their Table
1]{2010MNRAS.406..673M}: cluster members were defined as those with
the same friends-of-friends identification number and $M_{200}$ of
their parent halo, and clusters with less than five members were
rejected. Next, galaxies with $gri$ magnitudes brighter than the KiDS
limiting magnitudes (see Fig. \ref{Fig:DataQuality}) were selected: we
further considered only those clusters with at least 10 galaxies after
the latter selection.

\begin{figure}
   \centering
   \includegraphics[width=\columnwidth]{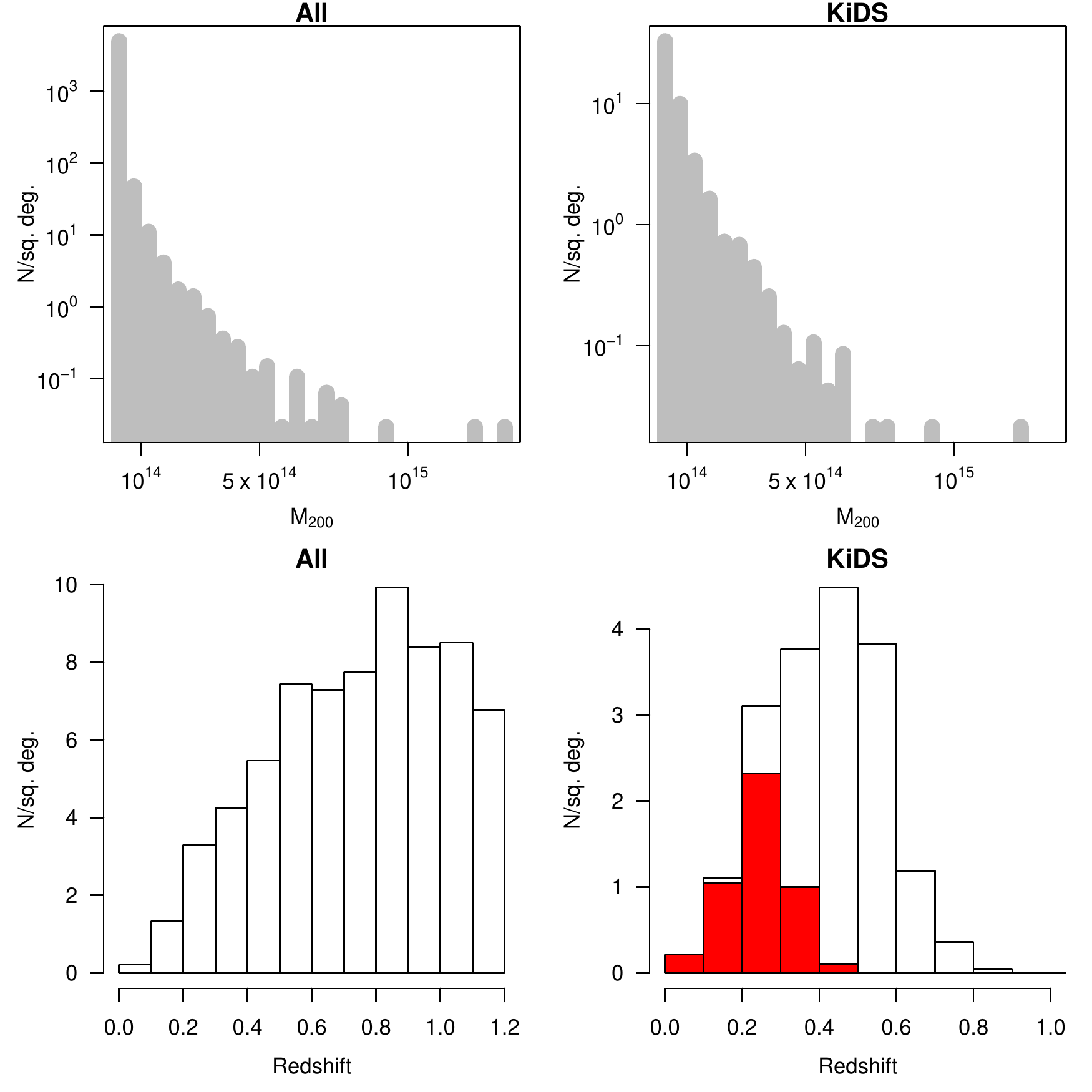}
   \caption{{\em Top:} Mass distribution of clusters in the H12 catalogues in the redshift range $0 < z <1$, before (left) and after (right) the selection based on the KiDS $gri$ limiting magnitudes. 
   {\em Bottom:} Redshift distribution; in the lower right panel, the results obtained using the SDSS magnitude limits are displayed for comparison (red histogram).  }
   \label{fig:clust_ms}
  \end{figure}

Figure~\ref{fig:clust_ms} shows
the mass and redshift distribution of the simulated clusters in H12
before ({\rm left}) and after ({\em right}) the KiDS magnitude cuts
were applied: it can be seen that a KiDS--like survey can probe galaxy
clusters in the $z \sim 0.4-0.8$ redshift range, extending the cluster
detection studies based on SDSS data, which are incomplete beyond $z
\sim 0.35$ \citep[e.g.][]{2014ApJ...785..104R, 2014arXiv1401.7716R}.
For comparison, in the lower right panel we overplot the redshift
distribution obtained adopting the SDSS limiting magnitudes ($i<21$
mag, $gr<22$ mag), showing the improvement expected with KiDS for $z >
0.4$. At the moment, a cluster detection analysis is on-going \citep{Radovich+15} using
the methods described in \citet{2011MNRAS.413.1145B}, which are based
on an optimal filter to find galaxy overdensities from the position,
photometry and possibly also photometric redshift of the galaxies in the
catalogues. Details and first results will be discussed in a separate
paper.

\subsection{High-redshift Quasar searches}
\label{Sec:HighZQSOs}

High redshift quasars are direct probes of the Universe less than 1 Gyr ($z > 5.7$) after the Big Bang. They provide fundamental constraints on the formation and growth of the first supermassive black holes (SMBHs), on early star formation, and on the chemical enrichment of the initially metal-free intergalactic and interstellar medium. 
The existing ensemble of high-redshift QSOs is dominated by luminous objects with exceptionally high accretion rates (close to Eddington) and very large SMBHs ($\sim10^9$ M$_\odot$). For several reasons it is needed to probe fainter QSOs, well below the "tip of the iceberg". This would allow to test SMBH early growth scenarios, which can predict a more common population of faint QSOs \citep{costa+14} with lower accretion rates. Furthermore, it would allow a better study of the symbiosis in growth of SMBHs and their stellar hosts over cosmic time. These are hampered currently by  potentially severe selection biases when comparing AGN-selected samples at high redshift to the host selected samples at low redshift \citep{willot+05,lauer+07}.

For these reasons we are building up a homogeneous sample of QSOs at $5.7<z<6.4$ by combining KiDS and VIKING and using the $i$-band drop-out technique. The 9-band $u$ through $K_s$ photometry from the combined surveys goes up to $\sim$2 mag deeper than SDSS, UKIDSS and the Panoramic Survey Telescope \& Rapid Response System 1 (Pan-STARRS1; \citealt{banados+14}). So far we have discovered nine such QSOs, where the first four are published in \cite{Venemans+15}. 
KiDS is also a useful ingredient in the detection of very high redshift ($z>6.4$) QSOs with VIKING. Adding the $i$-band KiDS data in the photometric selection  removes $\sim$50\% of these $Z$-band drop-out candidates \citep{venemans+13}.

\subsection{Strong gravitational lens searches}
\label{Sec:StrongLensing}

Strong gravitational lensing provides the most accurate and direct probe of mass in galaxies, groups and clusters of galaxies \citep{bolton+06,Tortora+10}.
The deep, subarcsecond seeing KiDS images are particularly suitable for a systematic census of
lenses based on the identification of arc-like structures
around massive galaxies, galaxy groups and galaxy
clusters.
The angular size of the Einstein ring $\theta_{\rm E}$ can be expressed as function of the velocity dispersion $\sigma_v$ as \citep{schneider92}:
\begin{equation}
\theta_{\rm E}\sim1''\times\frac{D_{\rm ds}}{D_{\rm s}}\frac{\sigma_v}{220 ~ \rm km s^{-1}},
\end{equation}
where $D_{\rm ds}$ and $D_{\rm s}$ are the angular diameter distance between the lens and source and to the source, respectively. For a typical FWHM$\sim0.7''$ in $r$-band, we can expect to detect
lensing arcs of gravitational structures with $\sigma_v> \rm 180~ kms^{-1}$. 

\begin{figure}
   \centering
   \includegraphics[width=8.5cm]{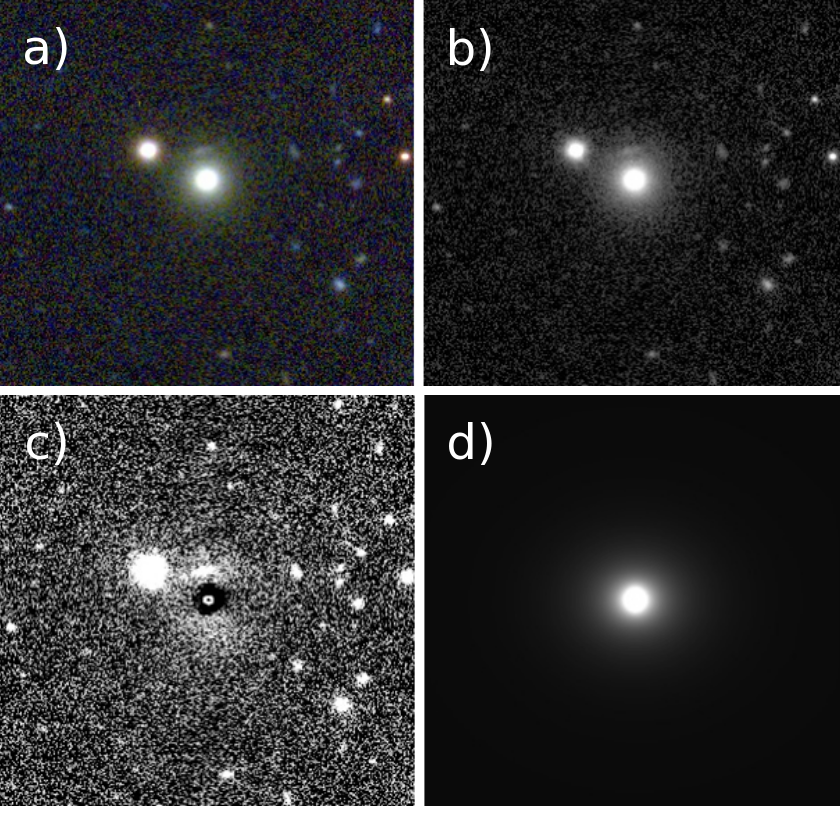}
   \caption{Example of strong lensing candidate identified in KiDS. {\em a)} colour image; {\em b)} $r$-band image; {\em c)} $r$-band image after subtraction of 2DPHOT galaxy model; {\em d)} 2DPHOT galaxy model. A faint blue arc is visible above the central galaxy and is clearly detected after model subtraction.}
   \label{fig:lensing}
  \end{figure}

A first search was based on visual inspection.
A sample of lens candidates was selected with the following two simple
selection criteria, using the photometric redshifts described in Sect. \ref{Sec:PhotzML} and the KiDS-ESO-DR1/2 $r$-band source lists. These criteria are aimed at maximising the lensing probability for
the KiDS photometric sensitivities: 1) $0.1 < z_{\rm {phot}} < 0.5$ and 2)
$r < 20$. Candidates are visually assessed by inspecting colour
images and $r$-band images where the galaxy model obtained with 2DPHOT
(see Sect. \ref{Sec:GalaxyParameters}) is subtracted. 
In a control set of
600 candidates in the area overlapping with SDSS, 18 potential lens candidates were identified, half of which have high significance \citep{Napolitano+15}. 
An example of a good lens candidate is shown in Fig. \ref{fig:lensing}.  
Another approach consists of automatic selection of lens candidates by filtering the KiDS-ESO-DR1/2 multi-band catalogues for massive early-type galaxies. This is done using colour-colour cuts and automated SED classifiers. In addition, candidates of strongly lensed quasars are identified based on colour and morphology. Candidates from both approaches have been selected further via visual inspection and are being followed up for spectral confirmation.

In the future, as the data volume grows, 
searches for strong lenses in large-area surveys such as KiDS will have to
be performed with (semi-)automated techniques
(\citealt{2006astro.ph..6757A}, \citealt{more12}, \citealt{gavazzi14},
\citealt{joseph14}), as the numbers of typical host galaxy candidates
can be of the order of thousands per square degree, rendering visual
inspection prohibitive. However, current automatic lens-finding tools
are not perfect and all automated techniques are still being tested on
simulated data (e.g. \citealt{metcalf14}, \citealt{gavazzi14}).

\section{Summary and outlook}
\label{Sec:Summary}

The Kilo-Degree Survey is a 1500 square degree optical imaging survey in four filters ($ugri$) with the VLT Survey Telescope. Together with its near-infrared sister survey VIKING, a nine-band $ugriZYJHK_s$ optical-infrared data set will be produced. While KiDS was primarily designed as a tomographic weak gravitational lensing survey, many secondary science cases are pursued.

In this paper the first two data releases (KiDS-ESO-DR1 and KiDS-ESO-DR2) are presented, comprising a total of 148 survey tiles, or $\sim$160 square degrees. The data products of these two  public data releases were produced using a fine-tuned version of the {\sc Astro-WISE} optical pipeline \citep{astrowise}, complemented with the automated {\sc Pulecenella} masking software and the {\sc KiDS-CAT} source extraction software. Data products include calibrated stacked images, weight maps, masks, source lists, and a multi-band source catalogue. Data can be accessed through the ESO Science Archive, the {\sc Astro-WISE} system, and the KiDS website (see Sect. \ref{Sec:DataAccess} for the relevant links).

The achieved data quality is very close to expectations, with a very small PSF ellipticity over the full FOV and typical limiting magnitudes (5$\sigma$ AB in 2\arcsec~aperture) of 24.3, 25.1, 24.9 and 23.8 in $u$, $g$, $r$ and $i$, respectively. As the designated lensing band, $r$-band receives the best dark seeing conditions, resulting in an median PSF size of $\sim$0.65\arcsec, with PSF size variations over the FOV usually $<$15\%. Within survey tiles, the photometry is stable to $\sim$2\%, but since the photometric calibration is performed using nightly zero-points, photometric offsets between tiles are present due to extinction variations; this will be improved in future data releases. The multi-band source catalogue provided is based on detection in $r$-band and 98\% complete to $r\simeq$ 24.0.

Early weak-lensing applications \citep{sifon15,viola15,vanuitert15} of KiDS data rely on the KiDS-ESO-DR1 and -DR2 data products presented here for photometric redshifts, while galaxy shape measurements are based on a dedicated pipeline \citep[see][]{erben+13,miller+13}. Other applications of KiDS-ESO-DR1 and -DR2 data include galaxy structural parameter studies, cluster detection, strong gravitational lens detection, and high-redshift QSO searches. The latter has resulted in the detection of nine $5.7<z<6.4$ QSOs to date (Venemans et al. in prep).

Future data releases, apart from adding more survey tiles to the currently covered area, are foreseen to include additional, value-added data products. During the first years of survey operations the observational data rate has been hampered by several factors, causing slower progress than anticipated. Together with ESO the KiDS team has been working on various improvements, which has led to a significant increase in survey speed since early 2014. At the current rate the survey is expected to be completed by 2019, but continuing efforts to enhance telescope and operational efficiency are expected to lead to further improvements.

\begin{acknowledgements}
Based on data products from observations made with ESO Telescopes at the La Silla Paranal Observatory under programme IDs 177.A-3016, 177.A-3017 and 177.A-3018, and on data products produced by Target/OmegaCEN, INAF-OACN, INAF-OAPD and the KiDS production team, on behalf of the KiDS consortium. The KiDS production team acknowledge support by NWO-M grants. OmegaCEN is financially supported by NOVA and Target. Members of INAF-OAPD and INAF-OACN also acknowledge the support from the Department of Physics \& Astronomy of the University of Padova, and of the Department of Physics of Univ. Federico II (Naples). Target is supported by Samenwerkingsverband Noord Nederland, European fund for regional development, Dutch Ministry of economic affairs, Pieken in de Delta, Provinces of Groningen and 
Drenthe. Target operates under the auspices of Sensor Universe. This work has made use of Astro-WISE, which is an on-going project which started from an EU FP5 RTD programme funded by the EC Action "Enhancing Access to Research Infrastructures". JdJ, EMH and NI are supported by NWO grant 614.061.610. MR acknowledges support from the Italian MIUR 2010-2011 through the PRIN "The dark Universe and the cosmic evolution of baryons: from current surveys to Euclid". AC and CH acknowledge support from the European Research Council under the EC FP7 grant number 240185. HHi is supported by the DFG Emmy Noether grant Hi 1495/2-1. GL wishes to acknowledge partial support from the Italian MIUR through the PRIN "Cosmology with Euclid". MP acknowledges financial support from PRIN-INAF 2014. CT has received funding from the European Union Seventh Framework Programme (FP7/2007-2013) under grant agreement n. 267251 ``Astronomy Fellowships in Italy'' (AstroFIt). JH is funded by NSERC. BJ acknowledges support by an STFC Ernest Rutherford Fellowship, grant reference ST/J004421/1. HH and MBE acknowledge support from the European Research Council FP7 grant number 279396. LVEK is supported in part through an NWO-VICI career grant (project number 639.043.308). MV is funded by grant 614.001.103 from the Netherlands Organisation for Scientific Research (NWO) and from the European Research Council under FP7 grant number 279396. This work was supported by the Deutsche Forschungsgemeinschaft via the project TR33 `The Dark Universe'.
\end{acknowledgements}

\clearpage

\onecolumn
\begin{appendix}

\section{Single-band source list columns}
\label{App:singleband}

The following table lists the columns that are present
in the single-band source lists provided in KiDS-ESO-DR1/2. Note that 
of the 27 aperture flux columns only the ones for the smallest aperture (2 pixels, or
0.4\arcsec\ diameter) and the largest aperture (200 pixels, or
40\arcsec\ diameter) are listed. Note: the label for the aperture of 28.5
pixels is FLUX\_APER\_28p5.

\begin{center}
\begin{longtable}{llll}
\caption{\label{Tab:singlebandcolumns} Columns provided in the single-band source lists.}\\
\hline\hline
Label & Format & Unit & Description \\
\hline
\endfirsthead

\multicolumn{4}{c}{\tablename\ \thetable{} -- continued from previous page}\\
\hline\hline
Label & Format & Unit & Description \\
\hline
\endhead

\hline
\multicolumn{4}{r}{{Continued on next page}}\\
\endfoot

\hline
\endlastfoot

2DPHOT & J & Source & classification (see section on star/galaxy separation)\\
X\_IMAGE & E & pixel & Object position along x      \\
Y\_IMAGE & E & pixel & Object position along y      \\
NUMBER & J &  & Running object number        \\
CLASS\_STAR & E &  & {\sc SExtractor} S/G classifier        \\
FLAGS & J &  & Extraction flags         \\
IMAFLAGS\_ISO & J &  & FLAG-image flags summed over the iso. profile\\
NIMAFLAG\_ISO & J &  & Number of flagged pixels entering      \\
IMAFLAGS\_ISO & J &  & Number of flagged pixels entering IMAFLAGS\_ISO\\
FLUX\_RADIUS & E & pixel & Fraction-of-light radii        \\
KRON\_RADIUS & E & pixel & Kron apertures in units of A or B  \\
FWHM\_IMAGE & E & pixel & FWHM assuming a gaussian core     \\
ISOAREA\_IMAGE & J & pixel$^2$ & Isophotal area above Analysis threshold     \\
ELLIPTICITY & E &  & 1 - B\_IMAGE/A\_IMAGE        \\
THETA\_IMAGE & E & deg & Position angle (CCW/x)       \\
MAG\_AUTO & E & mag & Kron-like elliptical aperture magnitude      \\
MAGERR\_AUTO & E & mag & RMS error for AUTO magnitude     \\
ALPHA\_J2000 & D & deg & Right ascension of barycenter (J2000)     \\
DELTA\_J2000 & D & deg & Declination of barycenter (J2000)      \\
FLUX\_APER\_2 & E & count & Flux vector within circular aperture of 2 pixels  \\
... & ... & ... & ...         \\
FLUX\_APER\_200 & E & count & Flux vector within circular aperture of 200 pixels   \\
FLUXERR\_APER\_2 & E & count & RMS error vector for flux within aperture of 2 pixels  \\
... & ... & ... & ...         \\
FLUXERR\_APER\_200 & E & count & RMS error vector for flux within aperture of 200 pixels  \\
MAG\_ISO & E & mag & Isophotal magnitude        \\
MAGERR\_ISO & E & mag & RMS error for isophotal magnitude     \\
MAG\_ISOCOR & E & mag & Corrected isophotal magnitude       \\
MAGERR\_ISOCOR & E & mag & RMS error for corrected isophotal magnitude    \\
MAG\_BEST & E & mag & Best of MAG\_AUTO and MAG\_ISOCOR     \\
MAGERR\_BEST & E & mag & RMS error for MAG\_BEST      \\
BACKGROUND & E & count & Background at centroid position      \\
THRESHOLD & E & count & Detection threshold above background      \\
MU\_THRESHOLD & E & arcsec$^{-2}$ & Detection threshold above background      \\
FLUX\_MAX & E & count & Peak flux above background      \\
MU\_MAX & E & arcsec$^{-2}$ & Peak surface brightness above background     \\
ISOAREA\_WORLD & E & deg$^2$ & Isophotal area above Analysis threshold     \\
XMIN\_IMAGE & J & pixel & Minimum x-coordinate among detected pixels     \\
YMIN\_IMAGE & J & pixel & Minimum y-coordinate among detected pixels     \\
XMAX\_IMAGE & J & pixel & Maximum x-coordinate among detected pixels     \\
YMAX\_IMAGE & J & pixel & Maximum y-coordinate among detected pixels     \\
X\_WORLD & D & deg & Baryleft position along world x axis    \\
Y\_WORLD & D & deg & Baryleft position along world y axis    \\
XWIN\_IMAGE & E & pixel & Windowed position estimate along x     \\
YWIN\_IMAGE & E & pixel & Windowed position estimate along y     \\
X2\_IMAGE & D & pixel$^2$ & Variance along x       \\
Y2\_IMAGE & D & pixel$^2$ & Variance along y       \\
XY\_IMAGE & D & pixel$^2$ & Covariance between x and y     \\
X2\_WORLD & E & deg$^2$ & Variance along X-WORLD (alpha)      \\
Y2\_WORLD & E & deg$^2$ & Variance along Y-WORLD (delta)      \\
XY\_WORLD & E & deg$^2$ & Covariance between X-WORLD and Y-WORLD     \\
CXX\_IMAGE & E & pixel$^{-2}$ & Cxx object ellipse parameter      \\
CYY\_IMAGE & E & pixel$^{-2}$ & Cyy object ellipse parameter      \\
CXY\_IMAGE & E & pixel$^{-2}$ & Cxy object ellipse parameter      \\
CXX\_WORLD & E & deg$^{-2}$ & Cxx object ellipse parameter (WORLD units)    \\
CYY\_WORLD & E & deg$^{-2}$ & Cyy object ellipse parameter (WORLD units)    \\
CXY\_WORLD & E & deg$^{-2}$ & Cxy object ellipse parameter (WORLD units)    \\
A\_IMAGE & D & pixel & Profile RMS along major axis     \\
B\_IMAGE & D & pixel & Profile RMS along minor axis     \\
A\_WORLD & E & deg & Profile RMS along major axis (WORLD units)   \\
B\_WORLD & E & deg & Profile RMS along minor axis (WORLD units)   \\
THETA\_WORLD & E & deg & Position angle (CCW/world-x)       \\
THETA\_J2000 & E & deg & Position angle (east of north) (J2000)    \\
ELONGATION & E & deg & A\_IMAGE/B\_IMAGE         \\
ERRX2\_IMAGE & E & pixel$^2$ & Variance of position along x     \\
ERRY2\_IMAGE & E & pixel$^2$ & Variance of position along y     \\
ERRXY\_IMAGE & E & pixel$^2$ & Covariance of position between x and y   \\
ERRX2\_WORLD & E & deg$^2$ & Variance of position along X-WORLD (alpha)    \\
ERRY2\_WORLD & E & deg$^2$ & Variance of position along Y-WORLD (delta)    \\
ERRXY\_WORLD & E & deg$^2$ & Covariance of position X-WORLD/Y-WORLD      \\
ERRCXX\_IMAGE & E & pixel$^{-2}$ & Cxx error ellipse parameter      \\
ERRCYY\_IMAGE & E & pixel$^{-2}$ & Cyy error ellipse parameter      \\
ERRCXY\_IMAGE & E & pixel$^{-2}$ & Cxy error ellipse parameter      \\
ERRCXX\_WORLD & E & deg$^{-2}$ & Cxx error ellipse parameter (WORLD units)    \\
ERRCYY\_WORLD & E & deg$^{-2}$ & Cyy error ellipse parameter (WORLD units)    \\
ERRCXY\_WORLD & E & deg$^{-2}$ & Cxy error ellipse parameter (WORLD units)    \\
ERRA\_IMAGE & E & pixel & RMS position error along major axis    \\
ERRB\_IMAGE & E & pixel & RMS position error along minor axis    \\
ERRA\_WORLD & E & deg & World RMS position error along major axis   \\
ERRB\_WORLD & E & deg & World RMS position error along minor axis   \\
ERRTHETA\_IMAGE & E & deg & Error ellipse position angle (CCW/x)     \\
ERRTHETA\_WORLD & E & deg & Error ellipse position angle (CCW/world-x)     \\
ERRTHETA\_J2000 & E & deg & J2000 error ellipse pos. angle (east of north)  \\
FWHM\_WORLD & E & deg & FWHM assuming a gaussian core     \\
ISO0 & J & pixel$^2$ & Isophotal area at level 0     \\
ISO1 & J & pixel$^2$ & Isophotal area at level 1     \\
ISO2 & J & pixel$^2$ & Isophotal area at level 2     \\
ISO3 & J & pixel$^2$ & Isophotal area at level 3     \\
ISO4 & J & pixel$^2$ & Isophotal area at level 4     \\
ISO5 & J & pixel$^2$ & Isophotal area at level 5     \\
ISO6 & J & pixel$^2$ & Isophotal area at level 6     \\
ISO7 & J & pixel$^2$ & Isophotal area at level 7     \\
SLID & K &  & Source list ID        \\
SID & K &  & Source ID within the source list     \\
HTM & K &  & Hierarchical Triangular Mesh (level 25)      \\
FLAG & K &  & Not used         \\
\hline
\end{longtable}
\end{center}

\newpage

\section{Multi-band catalogue}
\label{App:multiband}

The following table lists the columns that are present
in the multi-band catalog provided in KiDS-ESO-DR2. 

\begin{table}[hb]
\centering
\caption{Columns provided in the multi-band catalogue.}
\label{Tab:MultiBandColumns}
\begin{tabular}{l l l l}
\hline\hline
Label & Format & Unit & Description \\
\hline
\multicolumn{4}{c}{Measurements based on $r$-band detection image}\\
\hline
ID & 23A & & Source identifier \\
RAJ2000 & D & deg & Right ascension (J2000) \\
DECJ2000 & D & deg & Declination (J2000) \\
SG2DPHOT & K &  & Source classification  \\
A & D & pixel & Linear semi major axis   \\
B & D & pixel & Linear semi minor axis   \\
CLASS\_STAR & E &  & {\sc SExtractor} star/galaxy classifier   \\
KRON\_RADIUS & E & pixel & Kron-radius used for MAG\_AUTO  \\
POSANG & E & deg & Position angle  \\
SEQNR & K &  & $r$-band sequence number \\
\hline
\multicolumn{4}{c}{Measurements provided for each filter}\\
\hline
FLUXERR\_APER\_100\_<filter> & E & count & flux error in 100 pixel aperture \\
FLUXERR\_APER\_10\_<filter> & E & count & flux error in 10 pixel aperture \\
FLUXERR\_APER\_14\_<filter> & E & count & flux error in 14 pixel aperture \\
FLUXERR\_APER\_25\_<filter> & E & count & flux error in 25 pixel aperture \\
FLUXERR\_APER\_40\_<filter> & E & count & flux error in 40 pixel aperture \\
FLUXERR\_APER\_4\_<filter> & E & count & flux error in 4 pixel aperture \\
FLUXERR\_APER\_6\_<filter> & E & count & flux error in 6 pixel aperture \\
FLUXERR\_APERCOR\_100\_<filter> & E & count & corrected flux error in 100 pixel aperture \\
FLUXERR\_APERCOR\_10\_<filter> & E & count & corrected flux error in 10 pixel aperture \\
FLUXERR\_APERCOR\_14\_<filter> & E & count & corrected flux error in 14 pixel aperture \\
FLUXERR\_APERCOR\_25\_<filter> & E & count & corrected flux error in 25 pixel aperture \\
FLUXERR\_APERCOR\_40\_<filter> & E & count & corrected flux error in 40 pixel aperture \\
FLUXERR\_APERCOR\_4\_<filter> & E & count & corrected flux error in 4 pixel aperture \\
FLUXERR\_APERCOR\_6\_<filter> & E & count & corrected flux error in 6 pixel aperture \\
FLUX\_APER\_100\_<filter> & E & count & flux in 100 pixel aperture \\
FLUX\_APER\_10\_<filter> & E & count & flux in 10 pixel aperture \\
FLUX\_APER\_14\_<filter> & E & count & flux in 14 pixel aperture \\
FLUX\_APER\_25\_<filter> & E & count & flux in 25 pixel aperture \\
FLUX\_APER\_40\_<filter> & E & count & flux in 40 pixel aperture \\
FLUX\_APER\_4\_<filter> & E & count & flux in 4 pixel aperture \\
FLUX\_APER\_6\_<filter> & E & count & flux in 6 pixel aperture \\
FLUX\_APERCOR\_100\_<filter> & E & count & corrected flux in 100 pixel aperture \\
FLUX\_APERCOR\_10\_<filter> & E & count & corrected flux in 10 pixel aperture \\
FLUX\_APERCOR\_14\_<filter> & E & count & corrected flux in 14 pixel aperture \\
FLUX\_APERCOR\_25\_<filter> & E & count & corrected flux in 25 pixel aperture \\
FLUX\_APERCOR\_40\_<filter> & E & count & corrected flux in 40 pixel aperture \\
FLUX\_APERCOR\_4\_<filter> & E & count & corrected flux in 4 pixel aperture \\
FLUX\_APERCOR\_6\_<filter> & E & count & corrected flux in 6 pixel aperture \\
FLUX\_RADIUS\_<filter> & E & pixel & {\sc SExtractor} FLUX\_RADIUS  \\
FWHM\_IMAGE\_<filter> & E & pixel & {\sc SExtractor} FWHM\_IMAGE \\
FLAG\_<filter> & J & & {\sc SExtractor} extraction flag \\
IMAFLAGS\_ISO\_<filter> & J & & Mask flag \\
MAGERR\_AUTO\_<filter> & E & mag & RMS error for MAG\_AUTO  \\
MAGERR\_ISO\_<filter> & E & mag & RMS error for MAG\_ISO  \\
MAG\_AUTO\_<filter> & E & mag & Kron-like elliptical aperture magnitude \\
MAG\_ISO\_<filter> & E & mag & Isophotal magnitude   \\
NIMAFLAGS\_ISO\_<filter> & J &  & Number of masked pixels entering IMAFLAGS\_ISO  \\
ISOAREA\_IMAGE\_<filter> & J & pixel$^2$ & Isophotal aperture \\
XPOS\_<filter> & E & pixel & X pixel position <filter> coadd \\
YPOS\_<filter> & E & pixel & Y pixel position <filter> coadd \\
\hline
\end{tabular}
\end{table}

\end{appendix}

\end{document}